\documentclass[prper,aps,twocolumn,groupedaddress,floats,showpacs,final,superscriptaddress]{revtex4-1}
\usepackage[latin3]{inputenc}
\usepackage[makeroom]{cancel}
\usepackage{graphicx}
\usepackage{amsmath}
\usepackage{amsfonts}
\usepackage{amssymb}
\usepackage{color}
\usepackage[left=2cm,right=2cm,top=2cm,bottom=2cm]{geometry}
\usepackage[export]{adjustbox}
\usepackage{graphicx} 
\usepackage{dcolumn}
\usepackage{bm}
\usepackage{simplewick}
\usepackage{array}
\usepackage{appendix}
\RequirePackage[
   hyperindex,colorlinks,bookmarksnumbered,
   plainpages=true,pdfstartview=FitH]{hyperref}
\hypersetup{linkcolor=blue,urlcolor=blue,citecolor=blue}
\usepackage{hyperref}

\definecolor{purple}{rgb}{0.5,0,0.6}

\usepackage{ulem}
\renewcommand{\emph}[1]{\textit{#1}}
\definecolor{darkblue}{rgb}{0,0,0.5}
\definecolor{darkgreen}{rgb}{0,0.5,0}
\definecolor{darkred}{rgb}{.7,0,0}
\definecolor{purple}{rgb}{0.5,0,0.6}
\definecolor{orange}{rgb}{1,0.5,0}
\definecolor{grey}{rgb}{.6,.6,.6}
\definecolor{lightpink}{rgb}{1,0.7,0.75}
\definecolor{pink}{rgb}{1,0.4,0.58}
\definecolor{deeppink}{rgb}{1,0.08,0.58}

\newcommand{\DK}[1]{{\color{black} #1}}
\newcommand{\jvd}[1]{{\color{black}{#1}}} 
\newcommand{\cm}[1]{{\color{black}{#1}}} 

\newcommand{\pdag}{{\phantom{\dagger}}}
 
\renewcommand{\emph}[1]{\textit{#1}}

\newcommand{\imp}{{\rm imp}}
\newcommand{\bulk}{{\rm bulk}}

\newcommand{\eff}{{\rm eff}}

\begin{document}
\author{D. B. Karki}
\affiliation{The  Abdus  Salam  International  Centre  for  Theoretical  Physics  (ICTP),
Strada  Costiera 11, I-34151  Trieste,  Italy}
\affiliation{International School for Advanced Studies (SISSA), Via Bonomea 265, 34136 Trieste, Italy}
\author{Christophe Mora}
\affiliation{Laboratoire Pierre Aigrain, {\'E}cole
Normale Sup{\'e}rieure,
PSL Research University, CNRS, Universit{\'e} Pierre et Marie Curie,
Sorbonne Universit{\'e}s, Universit{\'e} Paris Diderot, Sorbonne Paris-Cit{\'e},
24 rue Lhomond, 75231 Paris Cedex 05, France}
\author{Jan von Delft}
\affiliation{Physics Department, Arnold Sommerfeld Center for Theoretical Physics and Center for NanoScience, Ludwig-Maximilians-Universit{\"a}t M{\"u}nchen, 80333 M{\"u}nchen, Germany}
\author{Mikhail N. Kiselev}
\affiliation{The  Abdus  Salam  International  Centre  for  Theoretical  Physics  (ICTP),
Strada  Costiera 11, I-34151  Trieste,  Italy}

\title{Two-color Fermi liquid theory for transport through a multilevel Kondo impurity}

\date{\today}

\begin{abstract}
\noindent We consider a quantum dot with ${\cal K}{\geq} 2$ orbital levels occupied by two electrons connected to two electric terminals. The 
generic model 
is given by a multi-level Anderson Hamiltonian. The weak-coupling theory at the particle-hole symmetric point is governed by a two-channel $S{=}1$ Kondo model characterized by intrinsic channels asymmetry. 
Based on a conformal field theory approach 
we derived an effective Hamiltonian at a strong-coupling fixed point. 
The Hamiltonian capturing the low-energy physics of a two-stage Kondo screening represents the quantum impurity by a two-color local Fermi-liquid. Using non-equilibrium (Keldysh) perturbation theory around the strong-coupling fixed point we analyse the transport properties of the model at finite temperature, Zeeman magnetic field and source-drain voltage
applied across the quantum dot. We compute the Fermi-liquid transport constants and discuss different universality classes associated with emergent symmetries. 
\end{abstract}


\maketitle

\section{Introduction}\label{Int}
It is almost four decades since the seminal work of Nozieres and Blandin 
(NB)~\cite{Nozieres_Blandin_JPhys_1980} about the Kondo effect in \textit{real} metals. The concept of the Kondo effect studied for impurity spin $S{=}1/2$ interacting with a single orbital channel ${\cal K}{=}1$ of conduction electrons \cite{Kondo, abrikosov_1965, shul_1965, Anderson_Yuval, Anderson_Yuval_Hamann, Migdal_Abrikosov, FZ_1971, Nozieres, AFFLECK_NPB_1990} was extended 
in \cite{Nozieres_Blandin_JPhys_1980} for arbitrary spin $S$ and arbitrary number of channels ${\cal K}$. A detailed classification
of possible ground states corresponding to the under-screened ${\cal K} {<} 2S$,
fully screened ${\cal K} {=} 2S$ and overscreened ${\cal K}{>} 2S$ Kondo effect has been given in \cite{wigmann_JETP(38)_1983, Andrei_RevModPhys_1983, Sacramento_CM(48)_1991, Cox_Adv_Phys(47)_1998}.
Furthermore,  it has been argued that in real metals the spin-$1/2$ single-channel Kondo effect is unlikely to be sufficient for the complete description of the physics of a magnetic impurity in a non-magnetic host. In many cases truncation of the impurity spectrum to one level is not possible and besides there are several orbitals of conduction-electrons which interact with the higher spin $S{>}1/2$ of the localized magnetic impurity \cite{Hewson}, giving rise to the phenomenon of multi-channel Kondo screening \citep{Affleck_Lud_PRB(48)_1993, Coleman_PRB(52)_1995}. In the fully screened case the conduction electrons completely screen the impurity spin to form a singlet ground state \cite{Andrei_PRL(52)_1984}.
As a result, the low-energy physics is described by a local Fermi-Liquid (FL) theory \cite{Nozieres, Nozieres_Blandin_JPhys_1980}. In the under-screened Kondo effect there exist not enough conducting channels to provide complete screening \cite{Coleman_PRL(94)_2005, Koller_Hewson_PRB(2005)}. Thus, there is a finite concentration of impurities with a residual spin contributing to the thermodynamic and transport properties. In contrast to the underscreened and fully-screened cases, the physics of the overscreened Kondo effect is not described by the FL paradigm resulting in dramatic change of the thermodynamic and transport behaviour \cite{Hewson}. 
\begin{figure}[h]
\begin{align}
\underbrace{\includegraphics[width=2.3cm, valign=c]{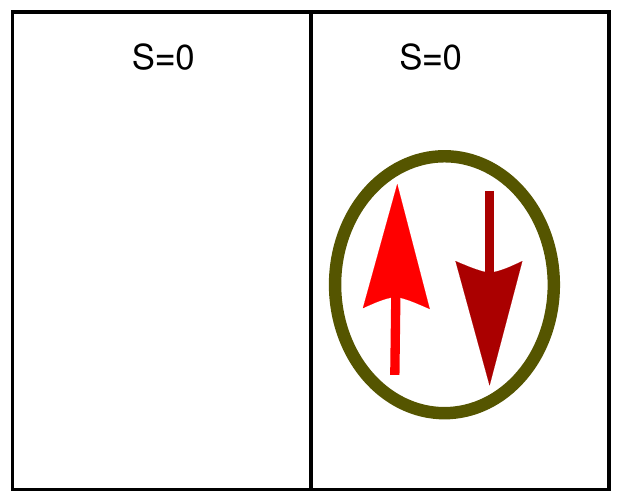}}_{\rm second\; stage}
\underbrace{\includegraphics[width=2.3cm, valign=c]{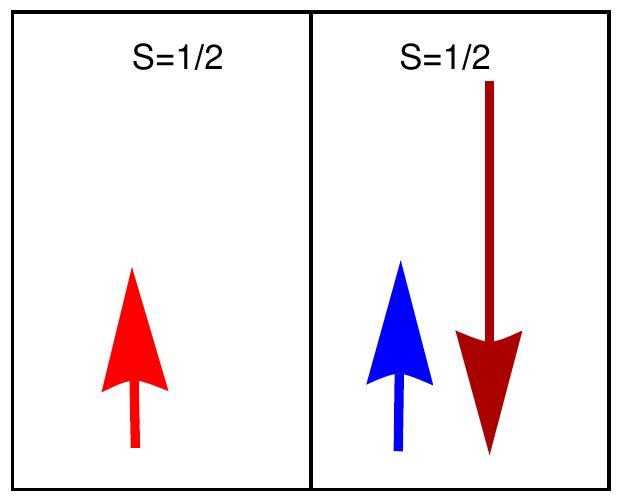}}_{\rm first\; stage}
\underbrace{\includegraphics[width=2.3cm, valign=c]{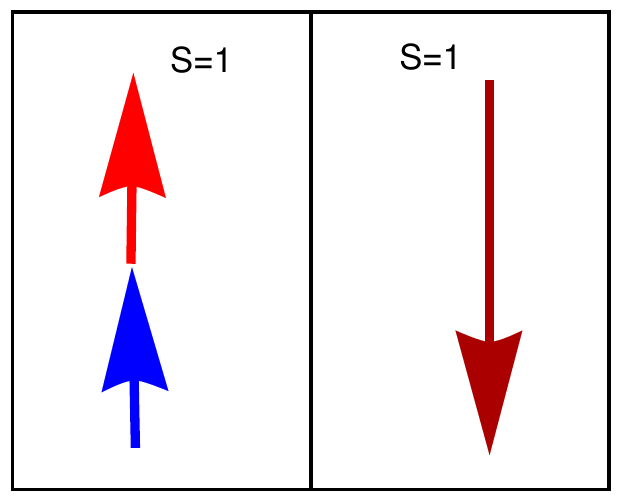}}_{\rm weak\; coupling}\nonumber\\
\includegraphics[width=7.5cm, valign=c]{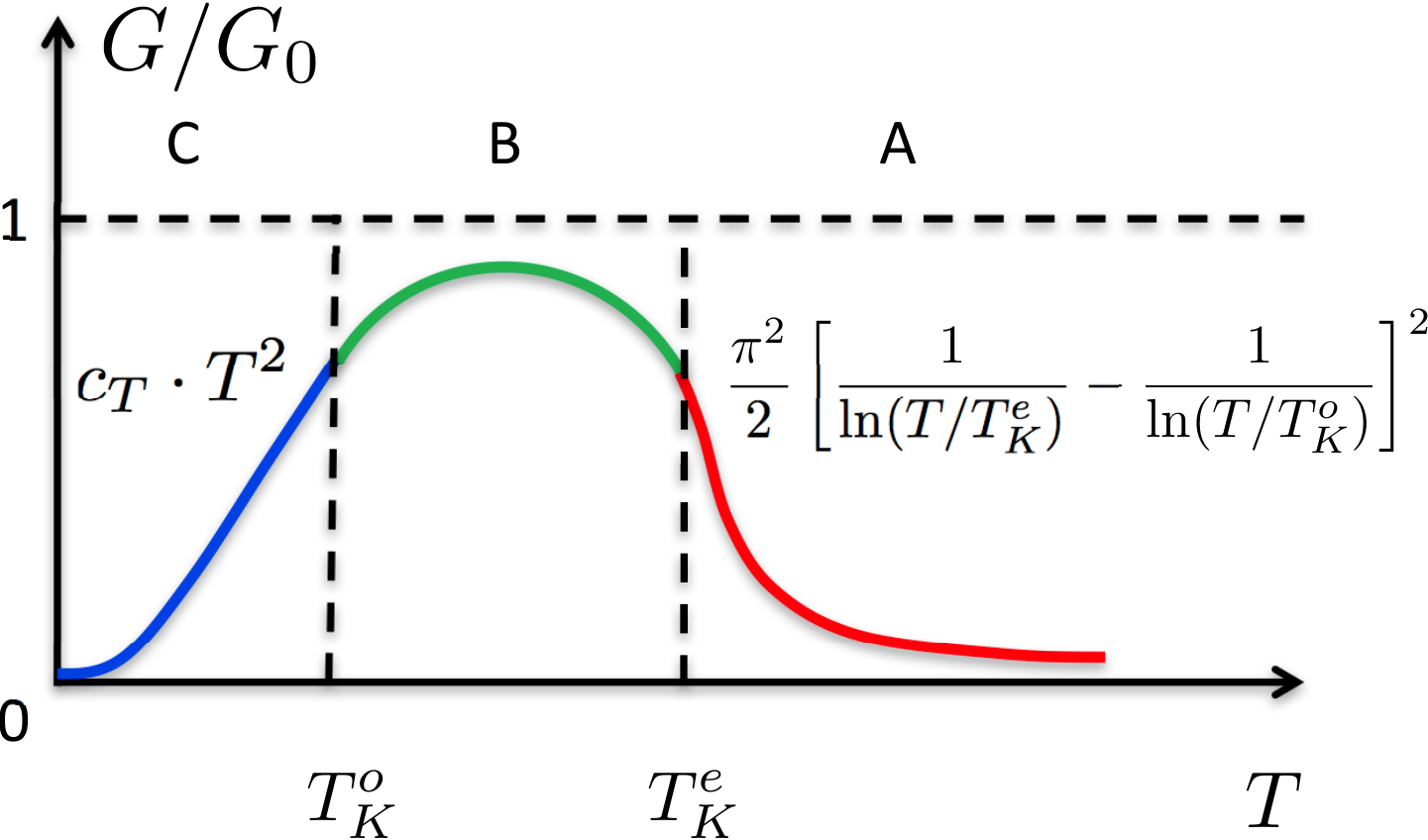}\nonumber
\end{align}
\caption{(Color online) Cartoon for non-monotonic behaviour of the
differential conductance $G/G_0$ ($G_0=2e^2/h$ is the conductance quantum)
as a function of temperature 
resulting from a two-stage Kondo effect. There are three characteristic regimes: (A) weak, (B) intermediate and (C) strong coupling.
Crossover energy scales $T_K^e$ and $T_K^o$ are defined in the Section \ref{mod}.
In the weak coupling (A)-regime the screening is absent (see top panel) and the transport coefficients are fully described by the perturbation theory \cite{GP_Review_2005}. In the intermediate regime (B), the Kondo impurity is partially screened (see the first stage at the top panel); the residual interaction of electrons with the under-screened spin {\color{black} is antiferromagnetic}
\cite{Nozieres_Blandin_JPhys_1980}.
The description of the FL transport coefficients in the strong coupling regime (C) at the second stage of the screening is the central result of the paper.}
\label{cart.2}
\end{figure}

The simplest realization of the multi-channel
fully screened Kondo effect is given by the model of a $S{=}1$ localized impurity screened by two conduction electron-channels. It has been predicted 
\cite{GP_Review_2005} that in spite of the FL universality class of the model, the transport properties of such FL are highly non-trivial. In particular, the screening develops in two stages (see Fig.~\ref{cart.2}), resulting in non-monotonic behaviour of the transport coefficients (see review~\cite{GP_Review_2005} for details).

The interest in the Kondo effect revived during the last two decades due to progress in fabrication of nano-structures \cite{revival}.
Usually in nanosized objects such as quantum dots (QDs), carbon nanotubes (CNTs), quantum point contacts (QPCs) etc., Kondo physics can be engineered by fine-tuning the external parameters (e.g. electric and magnetic fields) and develops  in the presence of several different channels of the conduction electrons coupled to the impurity.  Thus, it was timely \cite{Glazman_PRL_2001, GP_PRB_(64)_2001, revival, Hofstetter_Schoeller_2002,Pustilnik_Glazman_Hofstetter_2003,Hofstetter_Zarand_2004,GP_Review_2005} to uncover parallels between the Kondo physics in real metals and the Kondo effect in real quantum devices. The challenge of studying multi-channel Kondo physics \cite{Nozieres_Blandin_JPhys_1980, Affleck_Lud_PRB(48)_1993} was further revived in connection with possibilities 
to measure quantum transport in nano-structures experimentally \cite{Pierre_NAT(526)_2015,Keller_NAT(526)_2015,  Potok_NAT(446)_2007, Ralph_PRL(69)_1992, Ralph_PRB(51)_1995, Wiel_PRL(88)_2002} inspiring also many new theoretical suggestions \cite{Matveev_PRB(51)_1995, Cox_Adv_Phys(47)_1998, Rosch_PRL(87)_2001, Oreg_PRL(90)_2003, Coleman_PRL(94)_2005, Coleman_PRB(75)_2007,Kleeorin_Meir_2017}.

Unlike the $S{=}1/2$, ${\cal K}{=}1$ Kondo effect (1CK), the two-channel $S{=}1$ Kondo problem suffers from lack of universality for its observables \cite{Nozieres_Blandin_JPhys_1980}. The reason is that certain symmetries
(e.g. conformal symmetry) present in 1CK are generally absent in 
the two-channel $S{=}1$ model.
This creates a major obstacle for constructing a complete theoretical description in the low-energy sector of the problem. Such a description should, in particular, account for a consistent treatment
of the Kondo resonance \citep{Affleck_Lud_PRB(48)_1993} appearing in both orbital channels.
The interplay between two resonance phenomena, being the central reason for the non-monotonicity of transport coefficients \cite{GP_Review_2005}, has remained a challenging problem for many years \cite{Coleman_PRL(94)_2005, Coleman_PRB(75)_2007}. 

A sketch of the temperature dependence of the differential electric 
conductance is shown on Fig. \ref{cart.2}. The most intriguing result is that the differential conductance vanishes at both high and low temperatures, demonstrating the existence of two characteristic energy scales (see detailed discussion below). These two energy scales
are responsible for a two-stage screening of $S{=}1$ impurity. Following
\cite{Coleman_PRL(94)_2005, Coleman_PRB(75)_2007} we will refer to the
$S{=}1$, ${\cal K}{=}2$ Kondo phenomenon as the two-stage Kondo effect (2SK).

While both the weak (A) and intermediate (B) coupling regimes are well-described by the perturbation theory \cite{GP_Review_2005}, the  most challenging and intriguing question is the study of strong-coupling regime (C) where both  scattering channels are close to the resonance scattering. Indeed, the theoretical understanding of the regime C (in- and out-of-equilibrium) constitutes a long-standing problem that has remained open for more than a decade. Consequently, one would like to have a theory for the leading dependence of the {\color{black} electric current $I$ and} {\color{black} differential} conductance 
{\color{black} $G{=}\partial I/\partial V$} on magnetic field ($B$), temperature ($T$) and voltage ($V$), 
{\color{black}
$$G(B, T, V)/G_0=c_BB^2+c_T(\pi T)^2+c_VV^2.$$ 
Here $G_0{=}2e^2/h$ is unitary conductance.} Computation of these parameters $c_B$, $c_T$ and $c_V$ using a local FL theory and to show how are these related constitute the main message of this work.

In this paper we offer a full-fledged 
theory of the two-stage Kondo model at small but finite temperature, magnetic field and bias voltage to explain the charge transport (current, conductance) behaviour in the strong-coupling regime of the 2SK effect. The paper is organized as follows. In Section \ref{mod} we discuss the multi-level Anderson impurity model along with different coupling regimes. The FL-theory of the 2SK effect in the strong-coupling regime is addressed in Section \ref{flm}. We outline the current calculations which account for both elastic and inelastic effects using the non-equilibrium Keldysh formalism in Section \ref{cc}. In Section \ref{tp} we summarize our results for the FL coefficients in different regimes controlled by external parameters and discuss the universal limits of the theory. The Section \ref{diss} is devoted to discussing perspectives and open questions.
Mathematical details of our calculations are given in Appendices. 

\section{Model}\label{mod}
\begin{figure}[t]
\begin{align}
\includegraphics[width=7.8cm, valign=c]{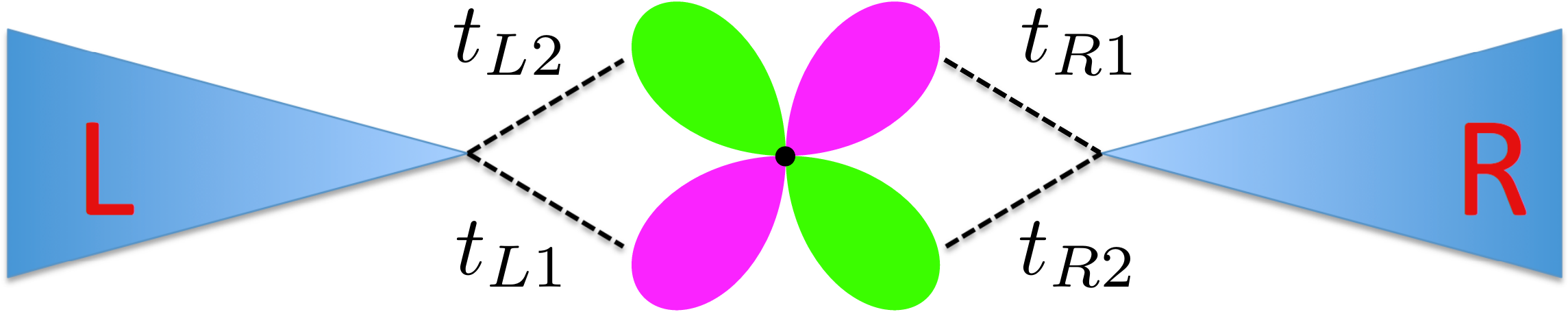}\nonumber\\
\underbrace{\includegraphics[width=3.9cm, valign=c]{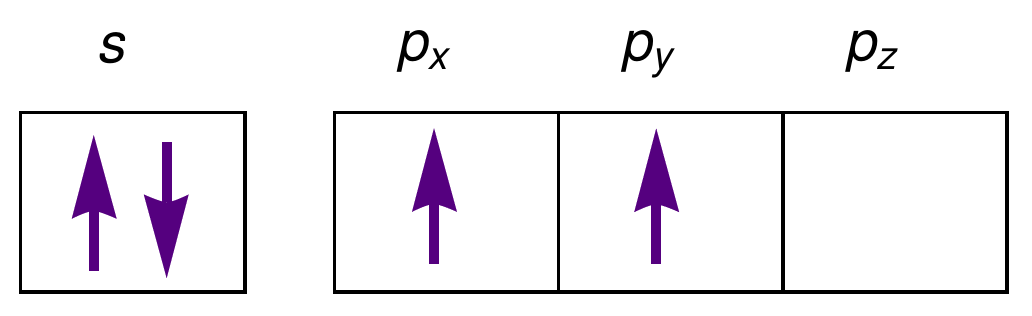}}_{\rm S=1}
\underbrace{\includegraphics[width=3.9cm, valign=c]{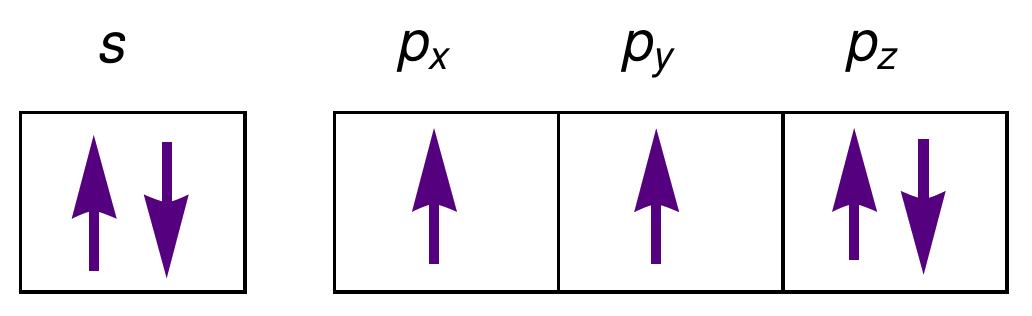}}_{\rm S=1}\nonumber
\end{align}
\caption{(Color online) 
Cartoon of some possible realization of a multi-orbital Anderson model setup: 
two degenerate $p$-orbitals (magenta and green) of a quantum dot 
are occupied by one electron each forming a triplet $S{=}1$ state in accordance with the Hund's rule (see lower panel). 
The third $p$-orbital (not shown) is either empty or doubly occupied. 
Two limiting cases are important: i) totally constructive interference
$t_{L1}{=}t_{L2}{=}t_{R1}{=}t_{R2}{=}t$; ii) totally destructive interference
$t_{L1}{=}t_{L2}{=}t_{R1}{=}t$, $t_{R2}{=}-t$. Besides, if 
$t_{L2}{=}t_{R2}{=}0$,
only one orbital is coupled to the leads, resulting in the 1CK model. If  
$t_{L2}{=}t_{R1}{=}0$, each orbital is coupled to a ``dedicated lead" and the net current through the dot is zero.}
\label{setup_1}
\end{figure}

We consider a multi-level quantum dot sandwiched between two external leads $\alpha$ (${=}L,R$) as shown in Fig.~\ref{setup_1}. The generic Hamiltonian is defined by the Anderson model
\begin{align}\label{model}
H&=\sum_{k\alpha\sigma}\left(\xi_k +\varepsilon_\sigma^Z\right) c^\dagger_{\alpha k \sigma}c_{\alpha k \sigma}+\sum_{\alpha k i \sigma} t_{\alpha i} c^\dagger_{\alpha k \sigma} d_{i \sigma} + \text{H.c.}\nonumber\\
&+\sum_{i\sigma}
(\varepsilon_i +\varepsilon_\sigma^Z) d^\dagger_{i \sigma}d_{i \sigma}
+E_c \hat{\cal N}^2- {\cal J} \hat{\mathbf{S}}^2,
\end{align}
where $c_\alpha$ stands for the Fermi-liquid quasiparticles of the source 
($L$) and the drain ($R$) leads, $\xi_k=\varepsilon_k-\mu$
is the energy of conduction electrons with respect to
the chemical potential $\mu$, 
and spin $\sigma=\uparrow{\color{black}(+)},\downarrow{\color{black}(-)}$
and $\varepsilon_\sigma^Z=-\sigma B/2$. The operator $d_{i\sigma}$ describes electrons with spin $\sigma$ in the $i$-th
orbital state of the quantum dot and $t_{\alpha i}$ are the tunneling matrix elements, as shown in Fig.~\ref{setup_1}. Here $\varepsilon_i+\varepsilon_\sigma^Z$ is the energy of the electron in $i$-th orbital
level of the dot in the presence of a Zeeman field $B$, $E_c$ is the charging energy (Hubbard interaction in the Coulomb blockade regime \cite{Matveev_PRB_1995}),
${\cal J}\ll E_c$ is an exchange integral accounting for Hund's rule \cite{Coleman_PRB(75)_2007} and $\hat{\cal N}=\sum_{i\sigma} d^{\dagger}_{i\sigma}d_{i\sigma}$ 
is the total number of electrons in the dot. We assume that the dot is occupied by two electrons, and thus the expectation value of $\hat{\cal N}$ is 
$\bar n_d=2$  and the total spin $S=1$ (see Fig.~\ref{setup_1}).
By applying a Schrieffer-Wolff (SW) transformation \cite{Schrieffer_Wolf_PR(149)_1966} to the Hamiltonian Eq.~\eqref{model} we exclude single-electron states in the dot and project out  
the effective Hamiltonian, written in the $L${-}$R$ basis, onto the spin-1 sector of the model:
\begin{equation}\label{effh}
H_{\eff}=\sum_{k\alpha\sigma}\xi_kc^{\dagger}_{\alpha k \sigma}c_{\alpha k \sigma}+\sum_{\alpha\alpha'}J_{\alpha\alpha'}\left[\textbf{s}_{\alpha\alpha'}\cdot\textbf{S}\right],
\end{equation}
with $\alpha,\alpha'=L, R$, $B{=}0$ and
\begin{align}
\mathbf{s}_{\alpha\alpha'} &= \frac{1}{2}\sum_{kk'\sigma_1\sigma_2}
c^\dagger_{\alpha k \sigma_1}\boldsymbol{\tau}_{\sigma_{12}}c_{\alpha' k'\sigma_2},\\ \mathbf{S}&=\frac{1}{2}\sum_{{i\sigma_1\sigma_2}}d^{\dagger}_{i\sigma_1} \boldsymbol{\tau}_{\sigma_{12}}d_{i\sigma_2},\\
J_{\alpha\alpha'}&=\frac{2}{E_c}\left(%
\begin{array}{cc}
  |t_{L1}|^2+ |t_{L2}|^2& t^*_{L2}t_{R2}+t^*_{L1}t_{R1} \\
  t_{L2}t^*_{R2}+t_{L1}t^*_{R1} & |t_{R2}|^2+ |t_{R1}|^2 \\
\end{array}%
\right),\label{exmat}
\end{align}
where we use the short-hand notation  $\boldsymbol{\tau}_{\sigma_{ij}}\equiv\boldsymbol{\tau}_{\sigma_{i}\sigma_j}$ for the Pauli matrices.

The determinant of the matrix $J_{\alpha\alpha'}$ in  Eq.~\eqref{exmat} is
non-zero provided that $t_{L2}t_{R1}{\neq} t_{L1}t_{R2}$. Therefore, one may assume without loss of generality that both eigenvalues of the matrix $J_{\alpha\alpha'}$ are non-zero and, hence, both scattering channels interact with the dot. {\color{black} There are, however, two important cases deserving an additional discussion.
The first limiting case is achieved when two eigenvalues of 
$J_{\alpha\alpha'}$ are equal and
the matrix $J_{\alpha\alpha'}$ is proportional to the unit matrix in any basis of electron states of the leads. As a result, the net current through impurity vanishes at any temperature, voltage and magnetic field \cite{Coleman_PRB(75)_2007}
(see Fig. \ref{cart.2}, {\color{black} showing that the differential conductance vanishes when symmetry between channels emerges}). This is due to destructive interference between two paths  \cite{Coleman_PRB(75)_2007}
(Fig. \ref{setup_1}) occurring when e.g. $t_{L1}{=}t_{L2}{=}t_{R1}{=}t$, $t_{R2}{=}-t$. {\color{black} Precise calculations done later in  the paper highlight the role of destructive interference effects and quantify 
how the current goes to zero in the vicinity of the symmetry point.} The second limiting case is associated with constructive interference between two paths (Fig. \ref{setup_1}) when $t_{L1}{=}t_{L2}{=}t_{R1}{=}t_{R2}{=}t$. In that case the determinant of the matrix $J_{\alpha\alpha'}$ in  Eq.~\eqref{exmat}  and thus also one of the eigenvalues of $J_{\alpha\alpha'}$, is zero. As a result, \cm{the} corresponding channel is completely decoupled from the impurity. The model then describes the under-screened $S{=}1$ single-channel Kondo effect.}

Applying the Glazman-Raikh rotation \cite{Glazman_Raikh_JETP_(47)_1988} 
$b^\dagger_{e/o}{=}(c^\dagger_L{\pm} c^\dagger_R)/\sqrt{2}$ to the effective Hamiltonian Eq.~\eqref{effh} we re-write the Kondo Hamiltonian in the diagonal basis \cite{symcop}, introducing two coupling constants $J_e$, $J_o$ 
\begin{align}\label{Heff}
\mathcal{H}_\eff&=\sum_a\left(H^a_0+J_{a} \mathbf{s}_{a}\cdot\mathbf{S}\right).
\end{align}

In writing Eq.~\eqref{Heff} we assigned the generalized index $``a"$ to represent the even and odd channels $(a{=}e,o)$. $H^a_0{=}\sum_{ak\sigma}(\varepsilon_k{-}\mu) b^\dagger_{ak \sigma}b^\pdag_{ak \sigma}$ is the non-interacting Hamiltonian of channel $a$ in the rotated basis. The spin density operators in the new basis are: $\mathbf{s}_a{=}1/2\sum_{kk'\sigma_1\sigma_2}b^{\dagger}_{ak\sigma_1} \boldsymbol{\tau}_{\sigma_{12}}b_{ak'\sigma_2}$. For equal leads-dot coupling, the $J_{a}$ are of the order of $t^2/E_c$. 
{\color{black}
The interaction between even and odd channels is generated by the next non-vanishing order of Schrieffer-Wolff transformation 
\begin{equation}
H_{eo}=-J_{eo} \mathbf{s}_{e}\cdot\mathbf{s}_o
\label{heo}
\end{equation}
where $J_{eo}$ is estimated as $J_{eo}{\sim}J_eJ_o/{\color{black}\max[E_c,\mu]}$.}
As a result this term is irrelevant in the weak coupling regime. However, we note that the sign of $J_{eo}$ is positive, indicating
the ferromagnetic coupling between channels necessary for the
complete screening of the $S=1$ impurity \cite{Nozieres_Blandin_JPhys_1980} (see Fig.~\ref{cart.2}).

 {\color{black} The Hamiltonian (\ref{Heff}) describes the
  weak coupling limit of the two-stage Kondo model. The coupling
  constants $J_e$ and $J_o$ flow to the strong coupling fixed point
  (see details of the renormalization group (RG) analysis \cite{Anderson,Migdal_Abrikosov,FZ_1971}
  in
  Appendix~\ref{weak_cop_reg}). In the leading-$\log$ (one loop RG)
  approximation, the two channels do not talk to each other.  
  {\color{black}As
    a result, two effective energy scales emerge, referred as Kondo
    temperatures, $T_K^a=D\exp(-1/(2N_F J_a))$ ($D$ is a bandwidth and
    $N_F$ is 3-dimensional electron's density of states in the leads).  These
    act as crossover energies, separating three regimes: the
    weak-coupling regime, $T\gg \max[T_K^a]$ (see Appendix
    ~\ref{weak_cop_reg}); the intermediate regime,
    $\min[T_K^a]\ll T \ll \max[T_K^a]$ {\color{black}
characterized by an incomplete screening (see Fig.~\ref{cart.2})
when one conduction channels (even) falls into a strong coupling regime
while the other channel (odd) still remains at the weak coupling}   
    (see Appendix
    ~\ref{int_cop_reg}); and the strong-coupling regime,
    $T\ll \min[T_K^a]$.} In the following section we discuss the
  description of the strong coupling regime by a local Fermi-liquid
  paradigm.  }

\section{Fermi-Liquid Hamiltonian}\label{flm}

{\color{black} The RG analysis of the Hamiltonian (\ref{Heff})
(see Appendix~\ref{weak_cop_reg} for details) shows that the 2SK model has \cm{a} unique strong coupling fixed point corresponding to
complete screening of the impurity spin.} This strong-coupling fixed point
is of the FL-universality class. In order to account
for existence of two different Kondo couplings in the odd and even
channels and the inter-channel interaction, we conjecture that
the strong-coupling fixed point Hamiltonian contains three leading
irrelevant operators:
\begin{equation}\label{lir}
H=-\sum_{aa'} \lambda_{aa'} 
:\! \mathbf{s}_a(0)\cdot \mathbf{s}_{a'}(0)\!:\;,
\end{equation}
with $\lambda_{ee} {=} \lambda_e$,
$\lambda_{oo} {=} \lambda_o$ and
$\lambda_{eo} {=} \lambda_{oe}$. The  notation  
$:...:$ corresponds  to a  normal  ordering  where  all  divergences
originating  from  bringing  two  spin  currents $\mathbf{s}_a$  close  to
each  other  are  subtracted.
The conjecture (\ref{lir}) is in the spirit of Affleck's ideas \cite{Affleck_Lud_PRB(48)_1993} of
defining leading irrelevant operators of minimal operator dimension
being simultaneously (i) local, (ii) independent of the impurity spin
operator $\mathbf{S}$, 
(iii) rotationally invariant and (iv) independent of
the local charge density. We do not assume any additional (SO(3) or SU(2)) symmetry in the channel subspace except at the symmetry-protected point 
$\lambda_e{=}\lambda_o{=}\lambda_{eo}{=}\lambda$. 
{\color{black} At this symmetry point a new conservation law for 
the total spin current \citep{Affleck_Lud_PRB(48)_1993} emerges and the Hamiltonian reads as
$$
H=-\lambda :\! \mathbf{S}(0)\cdot \mathbf{S}(0)\!:,\;\;\;\;\;
\mathbf{S}=\mathbf{s}_e+\mathbf{s}_o.
$$}
{\color{black} This symmetric point is obtained with the condition $J_e{=}J_o$ in $H_{\eff}$, see Eq.~\eqref{Heff}.}
{\color{black} Under this condition, as has been discussed in the previous section,} the net current through the impurity is zero 
due to totally destructive interference. This symmetry protects the zero-current state at any temperature, magnetic and/or electric field (see Fig.~\ref{setup_1}).

Applying the point-splitting procedure \cite{Affleck_Lud_PRB(48)_1993, HWDK_PRB_(89)_2014} to the Hamiltonian Eq.~\eqref{lir},  we get $H=H_e+H_o+H_{eo}$ with
\begin{align}\label{scoupH}
H_a&={-}\tfrac{3}{4} i\lambda_a   \sum_{\sigma} 
\left[ b^\dagger_{a\sigma} \tfrac{d}{dx}b^\pdag_{a\sigma}{-}
\left(\tfrac{d}{dx} b^\dagger_{a\sigma}\right)b^\pdag_{a\sigma}\right]
{+}\tfrac{3}{2}\lambda_a\rho_{a\uparrow}\rho_{a\downarrow},\nonumber\\
H_{eo}&=-\lambda_{eo}
\left[ :\!\mathbf{s}_e(0) \cdot \mathbf{s}_o(0)+\mathbf{s}_o(0) 
\cdot \mathbf{s}_e(0)\!:\right].
\end{align}
The Hamiltonian Eq.~\eqref{scoupH} accounts for two copies of the $s{=}1/2$ Kondo model at strong coupling with an additional ferromagnetic interaction between the channels providing complete screening at $T{=}0$.

{\color{black} An alternative derivation of the strong-coupling Hamiltonian~\eqref{scoupH} can be obtained, following Refs.~\cite{Mora_PRB_(80)_2009,Mora_Moca_Delft_Zarand_PRB(92)_2015,Filippone_PRB_(95)_2017}, with the most general form of the low-energy FL Hamiltonian. }
For the two-stage Kondo problem corresponding to the particle-hole symmetric limit of the two-orbital-level Anderson
model, \cm{it} is given by $H=H_0 + H_\alpha+H_\phi+H_\Phi$ with 
\begin{align}
H_0&{=}\phantom{-}\sum_{a\sigma}\int_\varepsilon \nu\left(\varepsilon + \varepsilon_\sigma^Z\right)
b^\dagger_{a\varepsilon\sigma}b^\pdag_{a\varepsilon\sigma}\nonumber\\
  H_\alpha& {=}{-}\sum_{a\sigma}
  \int_{\varepsilon_{1-2}} \frac{\alpha_{a}}{2 \pi}  
\left(\varepsilon_1+\varepsilon_2\right)
 \!  b^\dagger_{a\varepsilon_1\sigma}b^\pdag_{a\varepsilon_2\sigma}\! \nonumber
\\
H_\phi& {=}\phantom{-}\sum_{a} \int_{\varepsilon_{1-4}} \frac{\phi_{a}}{\pi\nu} 
: \! b^{\dagger}_{a\varepsilon_1\uparrow}b_{a\varepsilon_2\uparrow}
b^{\dagger}_{a\varepsilon_3\downarrow}b_{a\varepsilon_4\downarrow} \! :\nonumber
\\
H_\Phi&{{=}}{{-}}{\sum_{\sigma_{{1{-}4}}}}
{\int_{\varepsilon_{{1{{-}}4}}}}
{{\frac{\Phi}{{2}{\pi}{\nu}}}}
{{:}b^{{\dagger}}_{{o\varepsilon_{1}\sigma_{1}}}{\boldsymbol{\tau}_{{{\sigma_{12}}}}}
{b_{{o\varepsilon_{2}\sigma_{2}}}} {b^{{\dagger}}_{e\varepsilon_{3}\sigma_{3}}{\boldsymbol{\tau}_{\sigma_{34}}}
b_{e\varepsilon_{4}\sigma_{4}}{:}}}{,}
\label{HFL4}
\end{align}
where $\nu{=}1/(2\pi\hbar v_F) $ is the density of states per species for a one-dimensional channel. In Eq.~\eqref{HFL4} $H_{\alpha}$ describes energy-dependent elastic scattering 
\cite{Affleck_Lud_PRB(48)_1993}. The inter and intra-channel quasiparticle interactions responsible for the inelastic effects are described by $H_{\Phi}$ and $H_{\phi}$ respectively. The particle-hole symmetry of the problem forbids to have any second-generation of FL-parameters \cite{Mora_PRB_(80)_2009} in Eq.~\eqref{HFL4}. Therefore, the Hamiltonian Eq.~\eqref{HFL4} constitutes a
minimal model for the description of a local Fermi-liquid with two interacting resonance channels.
The direct comparison of the above FL-Hamiltonian with the strong-coupling Hamiltonian
Eq.~\eqref{scoupH} provides the relation between the FL-coefficients at PH symmetry, namely
$\alpha_a{=}\phi_a$ 
{\color{black} The Kondo floating argument (see \cite{Mora_PRB_(80)_2009}) recovers this relation.}
As a result we have three independent FL-coefficients 
$\alpha_e$, $\alpha_o$ and $\Phi$ which can be obtained from three independent measurements of the response functions. 
The FL-coefficients in Eq.~\eqref{HFL4} are related to the leading irrelevant coupling parameters $\lambda$'s in Eq.~\eqref{scoupH} as
\begin{align}
\alpha_a&=\phi_a=\frac{3\lambda_a\pi}{2}\;\;\text{and}\;\;\Phi=\pi\lambda_{eo},
\end{align}
{\color{black} The symmetry point $\lambda_e{=}\lambda_o{=}\lambda_{eo}{=}\lambda$
constrains $\alpha_e{=}\alpha_o{=}3\Phi/2$ in the Hamiltonian Eq.~\eqref{HFL4}.}

{\color{black} To fix three independent FL parameters in (\ref{HFL4}) in terms of physical observables, 
three equations are needed. 
Two equations are provided by specifying the spin susceptibilities of two orthogonal channels. The remaining necessary equation can be obtained by
considering the impurity contribution to specific heat.
It is proportional to an impurity-induced change in the total density of states per spin \cite{Hewson},
$\nu^\imp_{a\sigma}(\varepsilon){=}\frac{1}{\pi}\partial_\varepsilon
\delta^{a}_\sigma(\varepsilon)$, where $\delta^a_\sigma(\varepsilon)$ are energy  dependent scattering phases in odd and even channels (see the next Section  for more details) 
\begin{align}\label{cimp}
\frac{C^\imp}{C_\bulk} = \frac{\sum_{a\sigma} \frac{1}{\pi} 
\partial_\varepsilon \delta_\sigma^a (\varepsilon)|_{\varepsilon=0}}{4 \nu}
 = \frac{\alpha_e+\alpha_o}{2 \pi \nu} \; .
\end{align}
The quantum impurity contributions to the spin susceptibilities of the odd and even channels (see details in \cite{HWDK_PRB_(89)_2014}) are given by 
\begin{equation}\label{chimp}
\frac{\chi^{\text{imp}}_e}{\chi_{\text{bulk}}}= \frac{\alpha_e+\Phi/2}{\pi\nu},\;\;\;\;\;\;\;\;\;\;\;\;
\frac{\chi^{\text{imp}}_o}{\chi_{\text{bulk}}}= \frac{\alpha_o+\Phi/2}{\pi\nu}.
\end{equation}
The equations (\ref{cimp}-\ref{chimp}) fully determine three FL parameters
$\alpha_e$, $\alpha_o$ and $\Phi$ in (\ref{HFL4}). Total spin susceptibility 
$\chi^{\text{imp}}{=}\chi^{\text{imp}}_e{+}\chi^{\text{imp}}_o$ together with the impurity specific heat (\ref{cimp}) defines the Wilson ratio, 
$R{=}(\chi^{\text{imp}}/\chi_{\text{bulk}})/(C^{\text{imp}}/C_{\text{bulk}})$ 
\cite{Affleck_Lud_PRB(48)_1993}, \cite{Gogolin_book}  which measures the ratio of the total specific heat to the contribution originating from the spin degrees of freedom
\begin{align}\label{wr}
R=2\left[\frac{\alpha_e+\alpha_o+\Phi}{\alpha_e+\alpha_o}\right]=2\left[1 +\frac{2}{3} \frac{\lambda_{eo}}{\lambda_e+\lambda_o}\right].
\end{align}
For $\lambda_e{=}\lambda_o{=}\lambda_{eo}$, Eq.~\eqref{wr} reproduces
the value $R{=}8/3$ known for the two-channel, fully screened $S{=}1$
Kondo model \cite{AFFLECK_AP_1995}. 
If however $\lambda_{eo}{=}0$ we get $R{=}2$, in
agreement with the text-book result for two not necessarily identical
but independent replicas of the single channel Kondo model.
}

\section{Charge Current}\label{cc}
The current operator at position $x$ is expressed in terms of first-quantized operators $\psi$ attributed to the linear combinations of the Fermi operators in the leads
\begin{equation}\label{current}
\hat I(x){=} \frac{e\hbar}{2mi}\sum_\sigma\left[\psi^\dagger_\sigma(x)\partial_x \psi_\sigma (x) - \partial_x \psi^\dagger_\sigma(x)\psi_\sigma (x) \right].
\end{equation}
In the present case both types of quasi-particles $b_{ak\sigma} (a{=}e, o)$ interact with the dot. Besides, both scattering phases ($e/o$) are close to their resonance value $\delta^{e/o}_{0,\sigma}{=}\pi/2$. This is in striking contrast to
the single channel Kondo model, where one of the eigenvalues of the $2\times 2$ matrix of $J_{\alpha\alpha'}$ in Eq.~\eqref{exmat} is zero, and hence the corresponding degree of freedom is completely decoupled in the interacting regime. For the sake of simplicity, 
we are going to consider the 2SK problem
in the absence of an orbital magnetic field so that magnetic flux is zero. However, our results can be easily generalized for the case of finite orbital magnetic field. In this section we obtain an expression of charge current operator for the two-stage Kondo problem following the spirit of seminal works \cite{Mora_Leyronas_Regnault_PRL_(100)_2008, Mora_Clerk_Hur_PRB(80)_2009, Mora_Leyronas_Regnault_PRL_(100)_2008, Vitushinsky_Clerk_Hur_PRL_(100)_2008, Mora_PRB_(80)_2009, Mora_Schuricht_(89)_2014}. The principal idea behind the non-equilibrium calculations is to choose a basis of scattering states for the expansion of the current operator Eq.\eqref{current}. The scattering states in the first quantization representation are expressed as
\begin{equation}\label{even}\nonumber
\psi_{ek\sigma}(x){=}\frac{1}{\sqrt{2}}\begin{cases}
\left [e^{i(k_F+k) x} - S_{e,\sigma}(k)e^{-i(k_F+k) x}\right] &  x<0 \\
\left [e^{-i(k_F+k) x} - S_{e,\sigma}(k)e^{i(k_F+k) x}\right] & x>0
\end{cases}
\end{equation}

\begin{equation}\label{odd}\nonumber
\psi_{ok\sigma}(x){=}\frac{1}{\sqrt{2}}\begin{cases}
\left [\phantom{-}e^{i(k_F+k) x} - S_{o,\sigma}(k)e^{-i(k_F+k) x}\right] & x<0\\
\left [-e^{-i(k_F+k) x} + S_{o,\sigma}(k)e^{i(k_F+k) x}\right] & x>0
\end{cases}
\end{equation}
The phase shifts in even/odd channels are defined through the corresponding $S$-matrix via the relation $S_{a,\sigma}(k){=}e^{2i\delta^a_\sigma(\epsilon_k)}$. Proceeding to second quantization, we project the operator $\psi_{\sigma}(x)$ over the eigenstates $\psi_{ek\sigma}(x)$ and $\psi_{ok\sigma}(x)$, choosing $x<0$ far from the dot, to arrive at the expression

\begin{align}\label{wfn}
\psi_\sigma(x)= &\frac{1}{\sqrt{2}}\sum_{k \sigma} \Big[ (e^{i(k_F+k)} - S_{e,\sigma}(k)e^{-i(k_F+k)})b_{ek\sigma}\nonumber\\
&+(e^{i(k_F+k)} - S_{o,\sigma}(k)e^{-i(k_F+k)})b_{ok\sigma}\Big].
\end{align}
Substituting Eq.~\eqref{wfn} into Eq.~\eqref{current} and using $b_{a\sigma}(x){=}\sum_kb_{ak\sigma}e^{ikx}$ and $Sb_{a\sigma}(x){=}\sum_kS(k)b_{ak\sigma}e^{ikx}$, we obtain an expression for the  current for symmetrical dot-leads coupling,
\begin{figure}[h]
\begin{center}
\includegraphics[width=60mm]{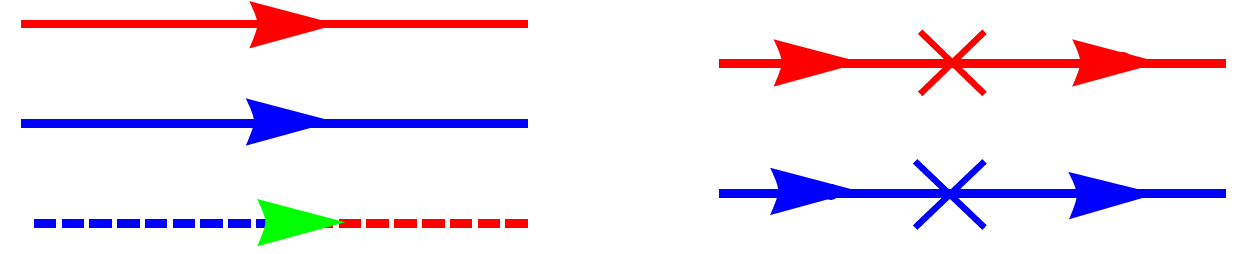}
\caption{(Color online) Left panel: Feynman codex used for the
  representation of different Greens functions: blue (red) line
  {\color{black} (in the black and white printout the colors are
    different by intensity of gray (red is more intensive))} for
  Green function of even (odd) channel $G_{e(o)}$ and the mixed line
  for the mixed Green function $G_{eo}$ {\color{black} (see definition
    in Section \ref{ccB1})}. Right panel: two-particle elastic
  vertices for even and odd channels. {\color{black} Crosses denote 
    energy-dependent scattering.}
}
\label{codexel}
\end{center} 
\end{figure}
\begin{equation}\label{finalcurr}
\hat I(x){=}{\frac{e}{2h\nu}}{\sum_\sigma}{\left[b^\dagger_{o\sigma}(x)b_{e\sigma}(x){-} b^{\dagger}_{o\sigma}({-}x) \mathcal{S} b_{e\sigma}({-}x)+ \text{H.c.}\right]},
\end{equation}
where $\mathcal{S}{=}S^\ast_oS_e$. There are two contributions to the charge current, coming from elastic and inelastic processes. The elastic effects are characterize by the energy-dependent phase-shifts, the inelastic ones are due to the interaction of Fermi-liquid  quasi-particles. In the following section we outline the elastic and inelastic current contribution of two-stage Kondo model Eq.~\eqref{HFL4}.
\subsection{Elastic current} 
We assume that the left and right scattering states are in thermal equilibrium at temperature $T_L{=}T_R{=}T$ and at the chemical potentials $\mu_R$ and $\mu_L{=}\mu_R{+}eV$. The population of states reads $2\langle b^\dagger_{ak\sigma} b_{ak'\sigma} \rangle{=}\delta_{kk'}\left[f_L(\varepsilon_k){+} f_R(\varepsilon_{k})\right]$ and $2\langle b^\dagger_{ak\sigma} b_{\bar{a}k'\sigma} \rangle{=}\delta_{kk'}\left[f_L(\varepsilon_k){-} f_R(\varepsilon_{k})\right]={}\delta_{kk'}\Delta f(\varepsilon_k)$ where $f_{L/R}(\varepsilon_{k}){=}f(\varepsilon_{k}{-}\mu_{L/R})$ and $f(\varepsilon_{k}){=}\left(1{+}\text{exp}\left[{\varepsilon_{k}}/{T}\right]\right)^{-1}$ is the Fermi-distribution function. {\color{black} The zero temperature conductance in the abscence of bias voltage is \cite{GP_Review_2005}}
{\color{black}
$$G(T=0,B\neq 0,V=0)/G_0 = B^2\left(\alpha_e-\alpha_o\right)^2.$$}
The elastic current in the absence of Zeeman field $B$ is the expectation value of the current operator Eq.~\eqref{finalcurr}. Taking the expectation value of Eq.~\eqref{finalcurr}
reproduces the Landauer-B\"uttiker equation \cite{Blanter_Nazarov}
 \begin{equation}\label{buttiker}
I_{\text{el}}=\frac{2 e}{h} \int^{\infty}_{-\infty} d\varepsilon T(\varepsilon)\Delta f(\varepsilon),
\end{equation}
where the energy dependent transmission coefficient, $T(\varepsilon){=}\frac{1}{2}\sum_\sigma\sin^2(\delta^e_\sigma(\varepsilon){-}\delta^o_\sigma(\varepsilon))$ and $\Delta f(\varepsilon){=}f_L(\varepsilon){-}f_R(\varepsilon)$. Diagrammatically (see Ref.~\cite{Affleck_Lud_PRB(48)_1993} and Ref.~\cite{HWDK_PRB_(89)_2014} for details), the elastic corrections to the current can be reabsorbed into a Taylor expansion 
for the energy-dependent phase shifts through the purely elastic contributions to quasi-particles self-energies \cite{Affleck_Lud_PRB(48)_1993}. That is the scattering phase-shifts can be read off \cite{Affleck_Lud_PRB(48)_1993} via the real part of the retarded self-energies $\Sigma^R_{a,\sigma}(\varepsilon)$ (see Fig. \ref{codexel}) as
\begin{equation}\label{delta_E}
\delta^a_\sigma(\varepsilon){=}-\pi\nu\text{Re}\Sigma^R_{a,\sigma}(\varepsilon)=\pi/2+\alpha_a\varepsilon.
\end{equation}
{\color{black} The Kondo temperatures of the two-channels 
in the  strong-coupling limit are defined as 
\begin{align}\label{strongTK}
T_K^a=\frac{1}{\alpha_a}.
\end{align}
This definition is consistent with Nozieres-Blandin
\cite{Nozieres_Blandin_JPhys_1980} and identical to that used in
\cite{HWDK_PRB_(89)_2014}, \DK{ however, is differ by the coefficient $\pi/4$ from the
    spin-susceptibility based definition
    \cite{Filippone_PRB_(95)_2017}.}
The elastic phase-shifts in the presence of the finite Zeeman field $B$ bears the form \cite{GP_Review_2005} (see schematic behaviour of 
$\delta^a_\downarrow(B) $ in Fig. \ref{cart.2a})
\begin{equation}\label{delta_B}
\delta^a_\sigma(B){=}\pi/2-(\alpha_a+\phi_a+\Phi)\bar\sigma B/2.
\end{equation}
\begin{figure}[h]
\begin{center}
 \includegraphics[width=50mm]{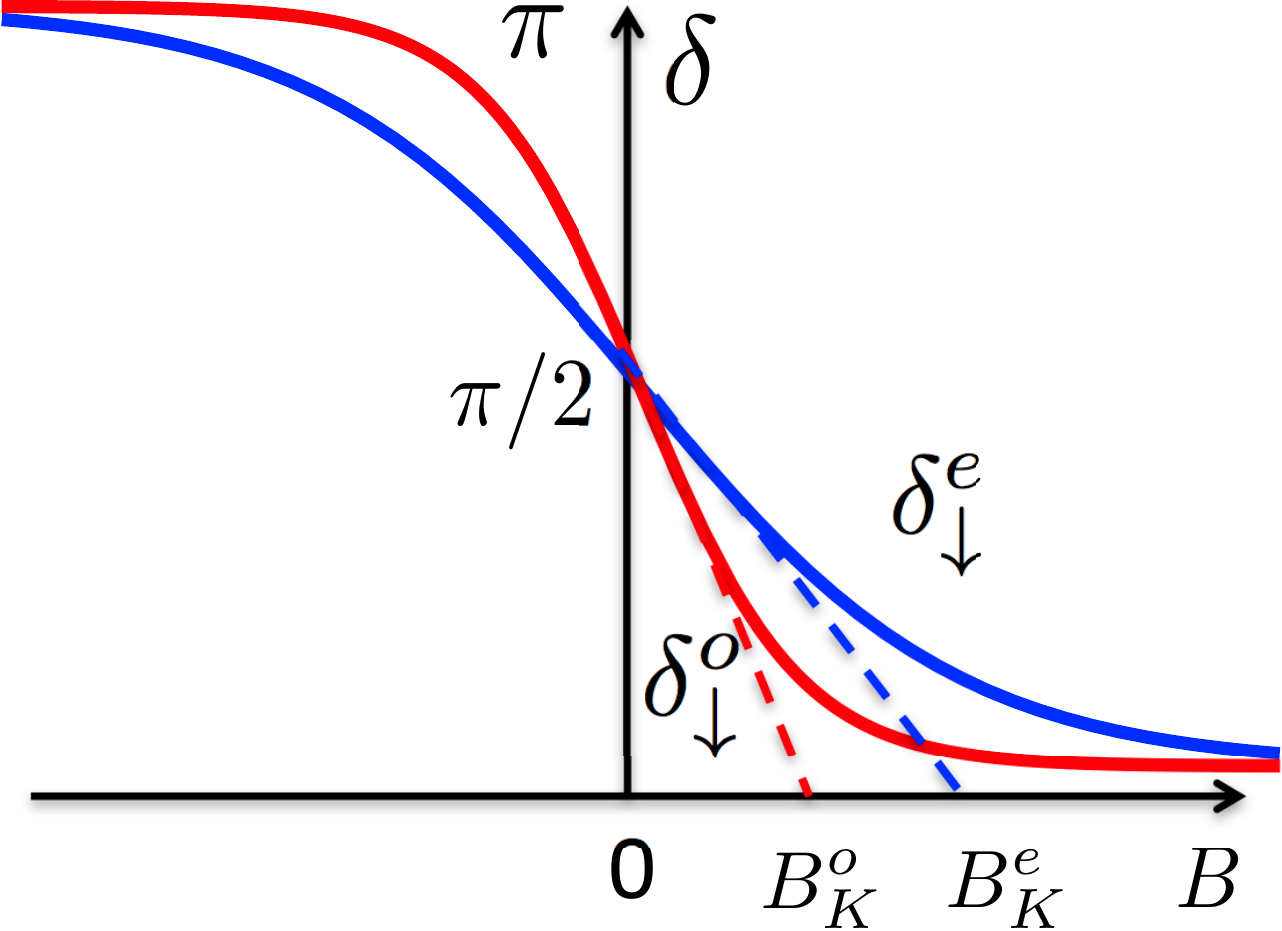}
 \end{center} 
\caption{(Color online) Schematic behaviour of the even (blue) and odd (red) scattering phases at $\sigma=\downarrow$ as a function of the Zeeman magnetic field. Both phases approach the resonance value $\pi/2$ at zero field. The tangential lines illustrate corresponding energy scales inversely proportional to the spin susceptibilities (\ref{chimp}) in the even/odd channels, {\color{black} $B_K^a=\pi/(2\alpha_a+\Phi)$} (see also Eqs.\eqref{delta_E}-\eqref{delta_B}).}
\label{cart.2a}
\end{figure}

Finally, we expand Eq.~\eqref{buttiker} up to second order in $\alpha_a$ to get the elastic contribution to the current \cite{Mora_Leyronas_Regnault_PRL_(100)_2008, Karki_MK_TE},
\begin{equation}\label{elastic}
\frac{I_{\text{el}}}{2e^2V/h}=\left[ B^2+ \frac{(eV)^2}{12}+ \frac{(\pi T)^2}{3}\right] (\alpha_e-\alpha_o)^2.
\end{equation}
\noindent The $B^2$ elastic term is attributed to the Zeeman field in Eq.~\eqref{model}. Note that
we do not consider the orbital effects assuming that the magnetic field is applied  parallel to the plane of the electron gas. The expression Eq.~\eqref{elastic} remarkably highlights the absence of a linear response at $T {=} 0$, $B{=} 0$, due to the vanishing of conductance when both scattering phases achieve the resonance value $\pi/2$.
The current is exactly zero at the symmetry point $\alpha_e{=}\alpha_o$ \cite{GP_Review_2005} due to the diagonal form of $S$-matrix characterized by two equal eigen values and therefore proportional to the unit matrix.
\subsection{Inelastic current}
To calculate the inelastic contribution to the current we apply the perturbation theory using Keldysh formalism \cite{Keldysh_SP_JETP(20)_1965},
\begin{equation}\label{wick}
\delta I_{\text{in}}=\langle T_C \hat{I}(t)e^{-i\int dt^{\prime} H_{\text{int}}(t^{\prime})}\rangle,
\end{equation}
\noindent where $H_{\text{int}}{=}H_{\phi}+H_{\Phi}$ and $C$ denotes the double-side 
$\eta{=}\pm$ Keldysh contour. Here $T_C$ is corresponding time-ordering operator. The average is performed with the Hamiltonian $H_0$.  The effects associated with quadratic Hamiltonian $H_\alpha$ are already accounted in $I_{\text{el}}$.
Therefore, to obtain the second-order correction to the inelastic current we proceed by considering $H_{\text{int}}{=}H_{\phi}+H_{\Phi}$, with the Feynman diagrammatic codex as shown in Fig.~\ref{codex}.

\begin{figure}[h]
\begin{center}
\includegraphics[scale=0.28]{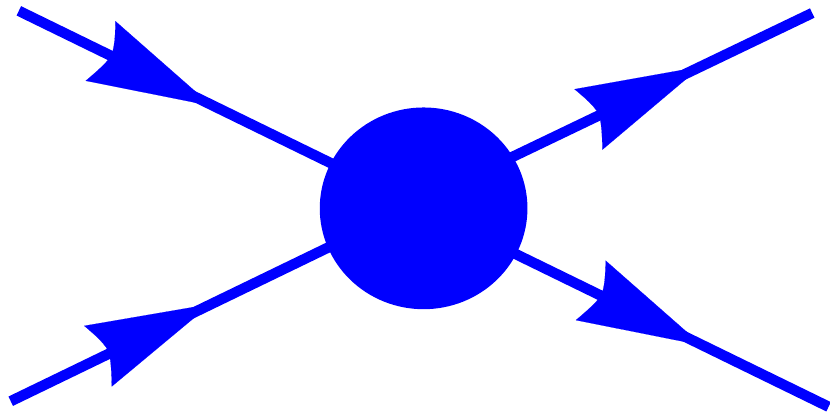}\;\;\includegraphics[scale=0.28]{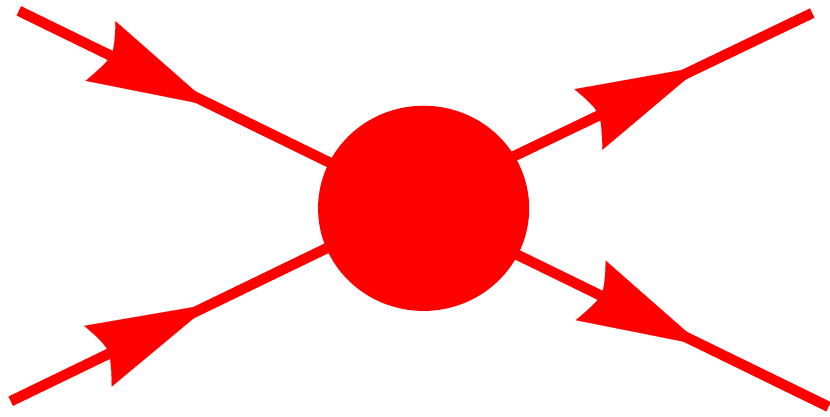}\;\;\includegraphics[scale=0.28]{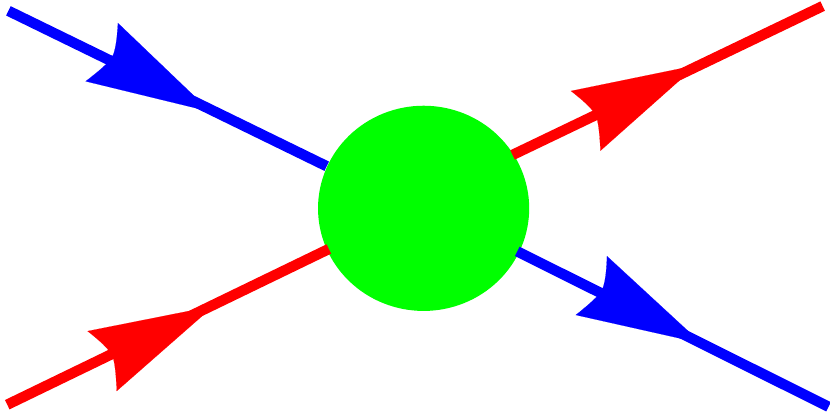}
\end{center}
\caption{(Color online) Feynman diagrammatic codex used for the calculation of inelastic current.
Blue (red) circles denote the density-density intra-channel interaction in even (odd) channel respectively
(see Eq.~\ref{HFL4}). Green circle denotes the inter-channel spin-spin interaction Eq.~\ref{HFL4}.
}\label{codex}
\end{figure}

The perturbative expansion of Eq.~\eqref{wick} in $(B,T,eV)\ll T_K^{\text{{\color{black} o}}}$ starts with the second-order contribution \cite{ Affleck_Lud_PRB(48)_1993} and is illustrated 
by Feynman diagrams of four types (see Fig.~\ref{type}).
The type-1 and type-2 diagrams contain only one mixed Green's function, GF (dashed line)
proportional to  $\Delta f(t){\sim} eV$, where  $\Delta f(t)$ is the Fourier transform of  $\Delta f(\varepsilon)$ defined in Eq.~\eqref{cu3}. Therefore, both diagrams fully define the linear-response contribution to the inelastic current, but also contain some non-linear
${\propto} (eV)^3$ contributions. The type-1 diagram contains the mixed GF
directly connected to the current vertex (Fig.~\ref{type}) and can be expressed in terms 
of single-particle self-energies. The type-2 diagram
contains the mixed GF completely detached from the current vertex and therefore can not
be absorbed into self-energies. We will refer to this topology of Feynman diagram as
a vertex correction. Note, that the second-order Feynman diagrams containing two (and also four) mixed GF are forbidden due to PH symmetry of the problem. The type-3 and type-4 diagrams contain three mixed GF's and therefore contribute only to the non-linear response being
proportional to $(eV)^3$. The type-3 diagram, similarly to the type-1 diagram, can be
absorbed into the single-particle self-energies. The type-4 diagram, similarly
to the type-2 diagram is contributing to the vertex corrections. This classification can be straightforwardly extended to higher order perturbation corrections for the current operator. Moreover, the diagrammatic series will have similar structure also for the Hamiltonians without particle-hole symmetry where more vertices are needed to account for 
different types of interactions. A similar classification can also be done for
current-current (noise) correlation functions \cite{Karki_MK_TBP}.
\begin{figure}[t]
\begin{center}
$\underbrace{
\includegraphics[height=22mm]{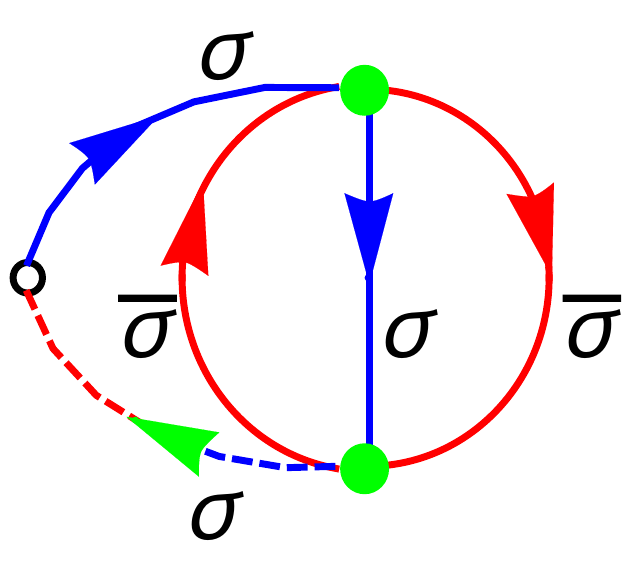}}_{\rm type\;1}\;\;\;
\underbrace{
\includegraphics[height=22mm]{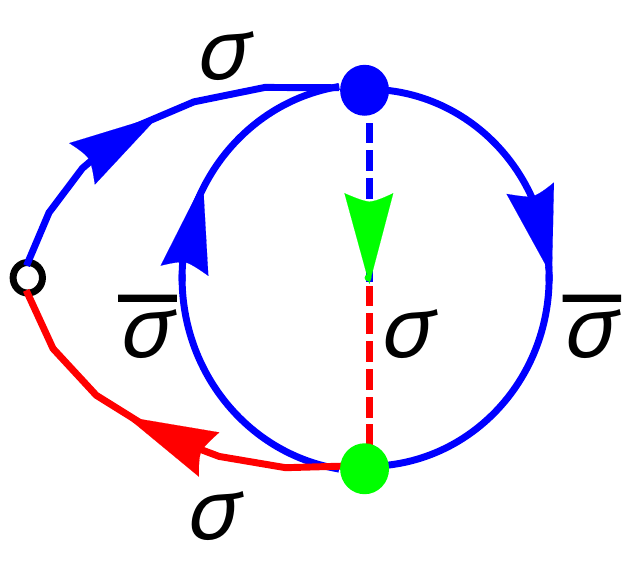}}_{\rm type\;2}
\linebreak
\underbrace{
\includegraphics[height=22mm]{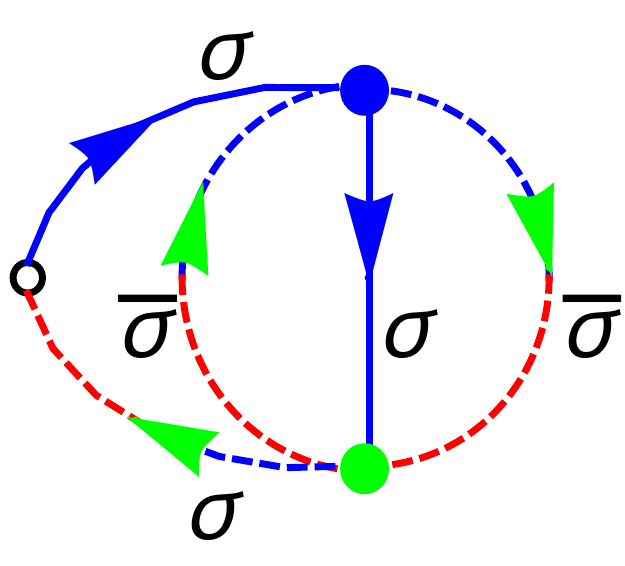}}_{\rm type\;3}\;\;\;
\underbrace{
\includegraphics[height=22mm]{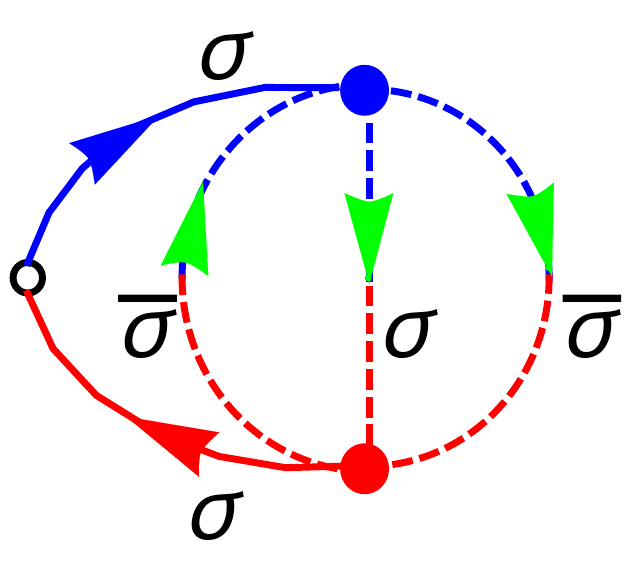}}_{\rm type\;4}$
\end{center}
\caption{(Color online) Examples of four different types of Feynamn diagrams contributing to the inelastic current. The open circle represents the current vertex. The other notations have been defined in Fig.~\ref{codexel} and Fig.~\ref{codex}.}\label{type}
\end{figure}
The mathematical details of the computation of the diagrammatic contribution of current correction diagrams type-1, type-2, type-3 and type-4 as shown in Fig.~\ref{type} proceed as follows:
\subsubsection{Evaluation of type-1 diagram}\label{ccB1}
The straightforward calculation of the Keldysh GFs at $x=0$ takes the form (see Refs.~\cite{Mora_Clerk_Hur_PRB(80)_2009,Karki_MK_TE} for details)
\begin{align}\label{matrixgf}
G_{aa}(\varepsilon)&=i\pi\nu\begin{bmatrix}
F_0(\varepsilon) &F_0(\varepsilon)+1\\
F_0(\varepsilon)-1& F_0(\varepsilon)
\end{bmatrix},\nonumber\\
G_{aa}(\varepsilon)&=i\pi\nu\begin{bmatrix}
1 &\phantom{-}1\\
1& \phantom{-}1
\end{bmatrix}\Delta f(\varepsilon),
\end{align}
where $F_0=f_L+f_R-1$ and we have neglected the principal part which does not contribute in the flat band model.
The current contribution proportional to $\Phi^2$ corresponding to the diagram of type-1 as shown in Fig.~\ref{type} is given by~\cite{Mora_Clerk_Hur_PRB(80)_2009}
\begin{equation}\label{first}
\delta I^{\Phi^2}_{\text{int}}=\frac{e}{\nu h}\sum_{\eta_1, \eta_2} \eta_1\eta_2\times \mathcal{Y}_1^{\eta_1, \eta_2},
\end{equation}
\noindent with
\begin{equation}\nonumber
{\mathcal{Y}_1^{\eta_1, \eta_2}}{=} {\int}{ \frac{d\varepsilon}{2\pi}}\left[{i\mathcal{S}}{G^{+\eta_1}_{ee}(-x, \varepsilon)}{\Sigma^{\eta_1\eta_2}(\varepsilon)}{G^{\eta_2-}_{eo}(x, \varepsilon)}{+}{\text{c.c.}}\right],
\end{equation}
\noindent where $\mathcal{S}=S^*_oS_e$, $\eta_{1/2}$ are the Keldysh branch indices which takes the value of $+$ or $-$. The self-energy $\Sigma^{\eta_1\eta_2}$ in real time is
\begin{align}\label{selfs}
{\Sigma^{\eta_1\eta_2}(t)}{=}{\left(\frac{\Phi}{\pi \nu^2}\right)^2}&{\sum_{k_1, k_2, k_3}} G^{\eta_1\eta_2}_{ee}(k_1, t)\nonumber\\
&\times {G^{\eta_2\eta_1}_{ee}(k_2, -t)} {G^{\eta_1\eta_2}_{ee}(k_3, t)}.
\end{align}
Using Eq.~\eqref{matrixgf} we express the diagonal and mixed GFs in real space as
\begin{align}\label{d1}
G^{\eta_1\eta_2}_{aa}(\alpha x, \varepsilon)&=i\pi\nu e^{i\alpha\varepsilon x/v_f}\left[F_0+\begin{cases}
\phantom{-}\eta_1 ,& \text{if } \alpha =\phantom{-} 1\\
-\eta_2,&  \text{if } \alpha= -1\\
\end{cases} \right], \nonumber \\
G^{\eta_1\eta_2}_{a\bar{a}}(x, \varepsilon)&=i\pi\nu e^{i\varepsilon x/v_f}\Delta f(\varepsilon),
\end{align}
The expression of corresponding GFs in real time is obtained by writing the Fourier transform of $\left(F_0(\varepsilon)\pm 1\right)$ as follows:
\begin{align}\label{c_c}
&\int \frac{d\varepsilon}{2\pi}\left(F_0(\varepsilon)\pm 1\right) e^{-i\varepsilon t}\nonumber\\
&=\frac{i}{2\pi}\left[\DK{\pm}\frac{\pi T}{\sinh(\pi T t)}\left(e^{-i\mu_L t}+e^{-i\mu_R t}\right)-2\frac{e^{\pm i D t}}{t}\right].
\end{align}
\noindent Summing Eq.~\eqref{first} over $\eta_1$ and $\eta_2$ using Eq.~\eqref{d1} results two terms involving $\Sigma^{++}-\Sigma^{--}$ and $\Sigma^{-+}-\Sigma^{+-}$. First term produces the contribution which is proportional to model cut-off $D$ is eliminated by introducing the counter terms in the Hamiltonian Eq.~\eqref{ct}. In rest of the calculation  we consider only the contribution which remain finite for $D\rightarrow \infty$. As a result we get
\begin{equation}\label{211}
\delta I^{\Phi^2}_{\text{int}}=\frac{2e\pi}{h} \int \frac{d\varepsilon}{2\pi}\left(\Sigma^{-+}(\varepsilon)-\Sigma^{+-}(\varepsilon)\right)i\pi\nu\Delta f(\varepsilon).
\end{equation}
\noindent In Eq.~\eqref{211} we used 
$\mathcal{S}+\mathcal{S}^*=2\cos(\delta^e_{0,\sigma}-\delta^o_{0,\sigma})=2$
with $\delta^e_{0,\sigma}{=}\delta^o_{0,\sigma}{=}\pi/2$. Fourier transformation of Eq.~\eqref{211} into real time takes the form
\begin{equation}\label{31s}
\delta I^{\Phi^2}_{\text{int}}=\frac{2e\pi}{h} \int dt \left(\Sigma^{-+}(t)-\Sigma^{+-}(t)\right)i\pi\nu\Delta f(-t).
\end{equation}
From Eq.~\eqref{c_c} the required Greens Function in real time are
\begin{equation}\label{gf1}
G^{+-}_{ee}(t)=-\pi\nu T \frac{\cos(\frac{eV}{2} t)}{\sinh(\pi T t)},
\end{equation}
\begin{equation}\label{gf2}
G_{eo}(t)=i\pi\nu T \frac{\sin(\frac{eV}{2} t)}{\sinh(\pi T t)},
\end{equation}
and $G^{-+}_{ee}(t)=G^{+-}_{ee}(-t)$. The self-energies in Eq.~\eqref{31s} are accessible by using above Greens functions Eq.~\eqref{gf1} and Eq.~\eqref{gf2} into self energy Eq.~\eqref{selfs}. Then Eq.~\eqref{31s} results in
\begin{equation}\label{t1}
\delta I^{\Phi^2}_{\text{int}}{=}\frac{2e\pi}{h}\left(\frac{\phi_e}{\pi \nu^2}\right)^2\times 2i(\pi\nu T)^4\int dt \frac{\cos^3(\frac{eV}{2} t) \sin(\frac{eV}{2} t)}{\sinh^4(\pi T t)}.
\end{equation}
The integral Eq.~\eqref{t1} is calculated in Appendix~\ref{integralcomp}. Hence the interaction correction to the current corresponding to the type-1 diagrams shown in Fig.~\ref{type} is
\begin{equation}
\frac{\delta I^{\Phi^2}_{\text{type}-1}}{2e^2V/h}{=}\left[A^{(1)}_V(eV)^2+A^{(1)}_T(\pi T)^2\right]\Phi^2,
\end{equation}
where $A^{(1)}_{V}=5/12\;\text{and}\;A^{(1)}_{T}=2/3$.
Alternatively, the calculation of the integral Eq.~\eqref{211} can be proceed by scattering T-matrix formalism. The single particle self energy difference accociated with the diagram of type-1 is expressed in terms of inelastic T-matrix to obtain \cite{Karki_MK_TE, GP_Review_2005}
\begin{equation}\label{sed}
\Sigma^{-+}(\varepsilon){-}\Sigma^{+-}(\varepsilon){=}\frac{\Phi^2}{i\pi \nu}\left[\frac{3}{4}(eV)^2{+}\varepsilon^2{+}(\pi T)^2\right].
\end{equation}
Using this self-energy difference and following the same way as we computed elastic current in Appendix \ref{ec}, one easily get the final expression for the current correction contributed by the diagram of type-1. 

\subsubsection{Evaluation of type-2 diagram}\label{ccB2}
\noindent The diagrammatic contribution of the type-2 diagram shown in Fig.~\ref{type} proportional to $\phi_e\Phi$ given by
\begin{equation}\label{vert11}
\delta I^{\phi_e\Phi}_{\text{int}}=\frac{e}{\nu h} \mathcal{J}=\frac{e}{\nu h}\sum_{\eta_1, \eta_2} \eta_1\eta_2 \mathcal{Y}_2^{\eta_1, \eta_2},
\end{equation}
\noindent with
\begin{equation}\label{j1}\nonumber
\mathcal{Y}_2^{\eta_1, \eta_2}{=}{\int}{\frac{d\varepsilon}{2\pi}}\left[{i\mathcal{S}}{G^{+\eta_1}_{ee}(-x, \varepsilon)}{\Lambda^{\eta_1\eta_2}_1( \varepsilon)}{G^{\eta_2-}_{oo}(x, \varepsilon)}{+\text{c.c.}}\right].
\end{equation}
\noindent The self energy part $\Lambda_1$ in real time is expressed as
\begin{align}\label{selfvert1}
\Lambda^{\eta_1\eta_2}_1(t)=\frac{\phi_e\Phi}{(\pi \nu^2)^2} &\sum_{k_1,k_2,k_3}G^{\eta_1\eta_2}_{ee}(k_1, t)\nonumber\\
&\times G^{\eta_2\eta_1}_{ee}(k_2, -t)G^{\eta_1\eta_2}_{eo}(k_3, t).
\end{align}

\noindent Now using Eq.~\eqref{d1} into Eq.~\eqref{vert11} followed by the summation over Keldysh indices, we get
\begin{align}\label{my1}
\mathcal{J}=2i\mathcal{S}(\pi\nu)^2 &\int dt[(F_0+1)(t)\Lambda^{-+}_1(-t)\nonumber\\
& -(F_0-1)(t)\Lambda^{+-}_1(-t)]+\text{c.c.}
\end{align}

\noindent Let us define the Greens function as $G^{+-/-+}_{ee}(t)=G^{+-/-+}_{oo}(t)\equiv G^{+-/-+}(t)$. Then we write
\begin{equation}\label{gft}
i\pi\nu (F_0\pm 1)(t)= G^{+-/-+}(t),
\end{equation} 
\noindent where $(F_0\pm 1)(t)$ is a shorthand notation for the Fourier transform of $F_0(\varepsilon)\pm 1$ defined by (\ref{c_c}). Hence, Eq.~\eqref{my1} takes the form
\begin{equation}\label{s1}
\mathcal{J}{=}{2\mathcal{S} \pi\nu}{\int}{dt}\left[{G^{+-}(t)}{\Lambda^{-+}_1({-t})}{-}{G^{-+}(t)}{\Lambda^{+-}_1({-t})}\right]{+\text{c.c.}}
\end{equation}
\noindent Now the self energies in Eq.~\eqref{selfvert1} cast the compact form 
\begin{align}
\Lambda^{-+}_1(-t)&=\frac{\phi_e\Phi}{(\pi \nu^2)^2}G^{-+}(-t)G^{+-}(t)G_{eo}(-t).\\
\Lambda^{+-}_1(-t)&=\frac{\phi_e\Phi}{(\pi \nu^2)^2}G^{+-}(-t)G^{-+}(t)G_{eo}(-t). 
\end{align}
\noindent Then the Eq.~\eqref{s1} becomes
\begin{equation}\label{de1}
\mathcal{J}= 4\mathcal{S} \pi\nu \frac{\phi_e\Phi}{(\pi \nu^2)^2}\int dt \left[G^{+-}(t)\right]^3G_{eo}(t)+\text{c.c.}
\end{equation}
\noindent Using the explicit expressions of the Greens functions Eqs.~\eqref{gf1} and ~\eqref{gf2} together with  Eq.~\eqref{de1} leads to
\begin{equation}\label{111}
\mathcal{J}= -4i(\pi\nu)^2\mathcal{S}T(\pi\nu T)^3 \frac{\phi_e\Phi}{(\pi \nu^2)^2}\int dt \frac{\cos^3(\frac{eV}{2}t) \sin(\frac{eV}{2}t)}{\sinh^4(\pi T t)}.
\end{equation}
Substituting the value of integral given by Eq.~\eqref{c9} into Eq.~\eqref{111} and using  Eq.~\eqref{vert11} we get
\begin{equation}
\frac{\delta I^{\phi_e\Phi}_{\text{type}-2}}{2e^2V/h}{=}\left[{A^{(2)}_V(eV)^2}{+}{A^{(2)}_T(\pi T)^2}\right]{\phi_e\Phi},
\end{equation}
where $A^{(2)}_{V}=-5/6\;\text{and}\; A^{(2)}_{T}=-4/3.$
\subsubsection{Evaluation of type-3 diagram}\label{ccB3}
\noindent Here we calculate the contribution to the current given by the diagram which consists of the self energy with two mixed Greens functions and one diagonal Greens function (type-3 diagram). The diagram shown in Fig.~\ref{type} describes correction  proportional to $\phi_e\Phi$ and is given by
\begin{equation}\label{type3}
\delta I^{\phi_e\Phi}_{\text{int}}=\frac{e}{\nu h}\sum_{\eta_1, \eta_2} \eta_1\eta_2\mathcal{Y}_3^{\eta_1, \eta_2},
\end{equation}
\noindent with
\begin{equation}\nonumber
{\mathcal{Y}_3^{\eta_1, \eta_2}}{=}{\int}{\frac{d\varepsilon}{2\pi}}\left[i\mathcal{S}G^{+\eta_1}_{ee}({-}x, \varepsilon)\Lambda^{\eta_1\eta_2}_2( \varepsilon)G^{\eta_2-}_{eo}(x, \varepsilon){+}{\text{c.c.}}\right].
\end{equation}
\noindent The self-energy $\Lambda^{\eta_1\eta_2}_2$ in real time is
\begin{align}\label{self}
\Lambda^{\eta_1\eta_2}_2(t)=\frac{\phi_e\Phi}{(\pi \nu^2)^2}&\sum_{k_1,k_2, k_3}G^{\eta_1\eta_2}_{eo}(k_1, t)\nonumber\\
&\times G^{\eta_2\eta_1}_{oe}(k_2, -t)G^{\eta_1\eta_2}_{ee}(k_3, t).
\end{align}
Summing Eq.~\eqref{type3} over $\eta_1$ and $\eta_2$ using Eq.~\eqref{d1}, we get
\begin{equation}\label{21}
\delta I^{\phi_e\Phi}_{\text{int}}{=}{\color{black}-}\frac{e}{\nu h} \times \pi\nu \mathcal{S} \int \frac{d\varepsilon}{2\pi}\left(\Lambda^{-+}_2(\varepsilon)-\Lambda^{+-}_2(\varepsilon)\right)i\pi\nu\Delta f(\varepsilon)+\text{c.c.}
\end{equation}
\noindent The Fourier transformation of Eq.~\eqref{21} into real time gives
\begin{equation}\label{31}
\delta I^{\phi_e\Phi}_{\text{int}}{=}{\color{black}-}\frac{e}{\nu h} \times \pi\nu \mathcal{S} \int dt \left(\Lambda^{-+}_2(t)-\Lambda^{+-}_2(t)\right)i\pi\nu\Delta f(-t) +\text{c.c.}
\end{equation}
\noindent Using the expressions of Greens functions in real time Eq.~\eqref{gf1} and Eq.~\eqref{gf2} allows to bring the interaction correction to the current Eq.~\eqref{31} 
to a compact form
\begin{equation}\label{41}
\begin{split}
\delta I^{\phi_e\Phi}_{\text{int}}{=}{\color{black}+}\frac{2e\pi}{h}{\times}2i(\pi\nu T)^4\frac{\phi_e\Phi}{(\pi\nu^2)^2}{\int} dt \frac{\cos(\frac{eV}{2}t) \sin^3(\frac{eV}{2}t)}{\sinh^4(\pi T t)}.
\end{split}
\end{equation}
Using Eq.~\eqref{c10} into Eq.~\eqref{41} we get
\begin{equation}\nonumber
\frac{\delta I^{\phi_e\Phi}_{\text{type}-3}}{2e^2V/h}{=}\left[A^{(3)}_V(eV)^2+A^{(3)}_T(\pi T)^2\right]\phi_e\Phi,
\end{equation}
where $A^{(3)}_{V}={\color{black}-}1/4\;\text{and}\; A^{(3)}_{T}=0.$
\subsubsection{Evaluation of type-4 diagram}\label{ccB4}
\noindent In this Section we calculate the diagrammatic contribution of the  $\phi_e\phi_o$ current diagrams (type-4 diagram) shown in Fig.~\ref{type}. Similar to type-2 diagram calculation, the current correction reads
\begin{equation}\label{vert1}
\delta I^{\phi_e\phi_o}_{\text{int}}=\frac{e}{\nu h} \mathcal{L}=\frac{e}{\nu h}\sum_{\eta_1, \eta_2} \eta_1\eta_2\mathcal{Y}_4^{\eta_1, \eta_2},
\end{equation}
\noindent with
\begin{equation}\label{j}
\mathcal{Y}_4^{\eta_1, \eta_2}{=}{\int} \frac{d\varepsilon}{2\pi}\left[i\mathcal{S}G^{+\eta_1}_{ee}(-x, \varepsilon)\Lambda^{\eta_1\eta_2}_3( \varepsilon)G^{\eta_2-}_{oo}(x, \varepsilon){+}\text{c.c.}\right].
\end{equation}
\noindent The self-energy part $\Lambda^{\eta_1\eta_2}_3$ is given by the expression
\begin{align}\label{selfvert}
\Lambda^{\eta_1\eta_2}_3(t)=\frac{\phi_e\phi_o}{(\pi \nu^2)^2}& \sum_{k_1, k_2, k_3}G^{\eta_1\eta_2}_{oe}(k_1, t)\nonumber\\
&\times G^{\eta_2\eta_1}_{eo}(k_2, -t)G^{\eta_1\eta_2}_{eo}(k_3, t).
\end{align}

\noindent Now substituting Eq.~\eqref{d1} into Eq.~\eqref{j} followed by the summation over Keldysh indices, we get
\begin{align}\label{my}
\mathcal{L}=2i\mathcal{S}(\pi\nu)^2 &\int dt[(F_0+1)(t)\Lambda^{-+}_3(-t)\nonumber\\
&-(F_0-1)(t)\Lambda^{+-}_3(-t)]+\text{c.c.}
\end{align}

\noindent Plugging in  Eq.~\eqref{gft} into Eq.~\eqref{my} results
\begin{equation}\label{s}
\mathcal{L}{=} 2\mathcal{S} \pi\nu{\int}{dt}\left[G^{+-}(t)\Lambda^{-+}_3({-}t){-}G^{-+}(t)\Lambda^{+-}_3({-}t)\right]{+}\text{c.c.}
\end{equation}
\noindent The self-energy Eq.~\eqref{s} takes the form
\begin{equation}
\Lambda^{-+}_3(-t)=\frac{\phi_e\phi_o}{(\pi \nu^2)^2}\left[G_{eo}(t)\right]^3=\Lambda^{+-}_3(-t).
\end{equation}
\noindent Hence combining Eq.~\eqref{gf1} and Eq.~\eqref{gf2} we bring the required integral Eq.~\eqref{s} to the form
\begin{equation}\label{11}
\mathcal{L}= {\color{black}-}\frac{\phi_e\phi_o}{(\pi \nu^2)^2} \times 4i\mathcal{S}\pi\nu(\pi\nu T)^4 \int dt \frac{\cos(\frac{eV}{2}t) \sin^3(\frac{eV}{2}t)}{\sinh^4(\pi T t)}+\text{c.c.}
\end{equation}
\noindent The integral in Eq.~\eqref{11} is given by Eq.\eqref{c10}. Hence plugging in Eq.~\eqref{11} into Eq.~\eqref{vert1} we obtain the current correction:
\begin{equation}
\frac{\delta I^{\phi_e\phi_o}_{\text{type}-4}}{2e^2V/h}{=}\left[A^{(4)}_V(eV)^2+A^{(4)}_T(\pi T)^2\right]\phi_e\phi_o,
\end{equation}
where $A^{(4)}_{V}={\color{black}+}1/2\;\text{and}\;A^{(4)}_{T}=0$.

As we discussed above, all the current diagrams are of the form of type-1, type-2, type-3 and type-4. However, same type of diagrams may contain different numbers of fermionic loops and also different spin combinations. In addition, there is the renormalization factor of 
$-\frac{1}{2}$ in $H_{\Phi}$, which has to be accounted for the diagrams containing at least one $\Phi$ vertex. Same type of diagrams containing at least one $\Phi$ vertex with different spin combination have the different weight factor because of product of Pauli matrices in $H_{\Phi}$. Each fermionic loop in the diagrams results in extra $(-1)$ multiplier in the corresponding weight factor. These facts will be accounted for by assigning the weight to the given current diagram (e.g. as shown in Fig.~\ref{weightfactor}, Fig.~\ref{weightfactor1} and Fig.~\ref{weightfactor2}).
\begin{figure}
\includegraphics[width=2.8cm, valign=c]{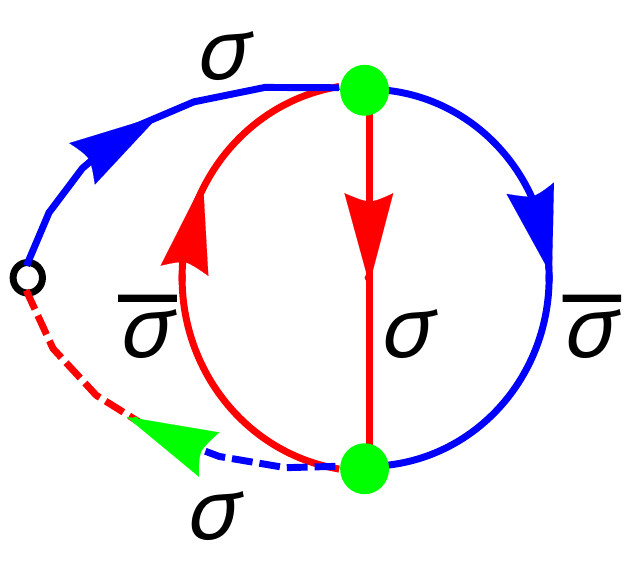}\;\;
\includegraphics[width=2.8cm, valign=c]{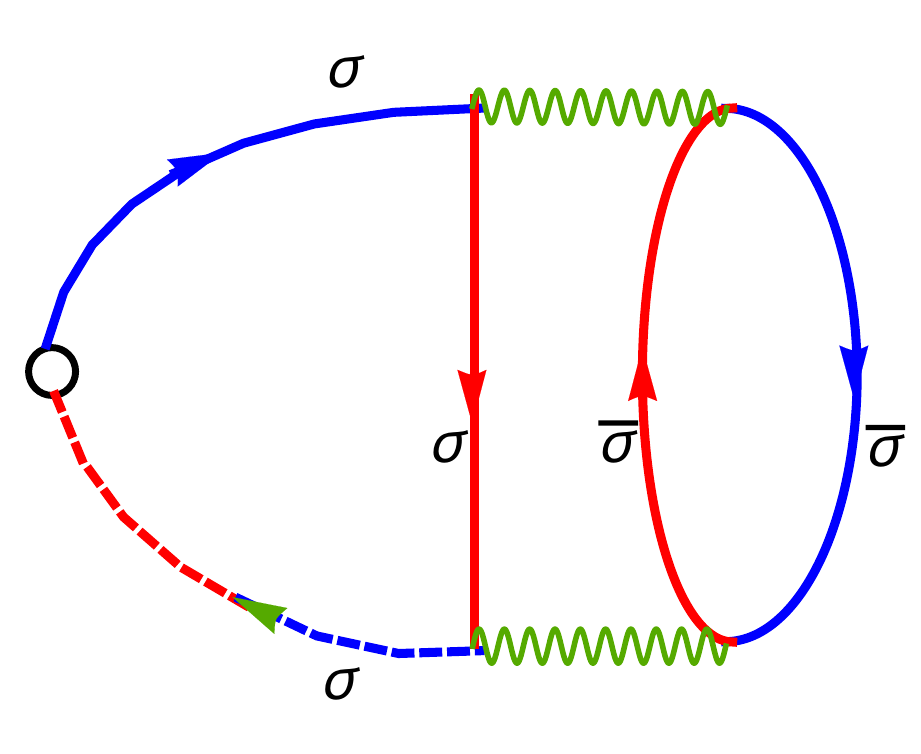}
\caption{(Color online) The $\Phi^2$ type-1 diagram (Left panel) and the corresponding diagram with the splitting of local $\Phi$ vertices (Right panel). In the diagram the upper $\Phi$ vertex consist in the Pauli matrices product $\tau_{\sigma\bar{\sigma}}.\tau_{\bar{\sigma}\sigma}{=}2$. Similarly the lower $\Phi$ vertex contain the product of $\tau_{\bar{\sigma}\sigma}.\tau_{\sigma\bar{\sigma}}{=}2$. The diagram contains the even number of fermionic loops (two) and hence no extra negative sing occur due to the fermionic loop. Each $\Phi$ vertex has the renormalization factor of $-\frac{1}{2}$. Hence the overall weight factor of this diagram is 
$\frac{1}{4}{\times} 4$ as will be seen in Fig.~\ref{LRFIG} and Fig.~\ref{BLRFIG}.}\label{weightfactor}
\end{figure}
\begin{figure}[b]
\includegraphics[width=2.8cm, valign=c]{y11.pdf}\;\;
\includegraphics[width=2.8cm, valign=c]{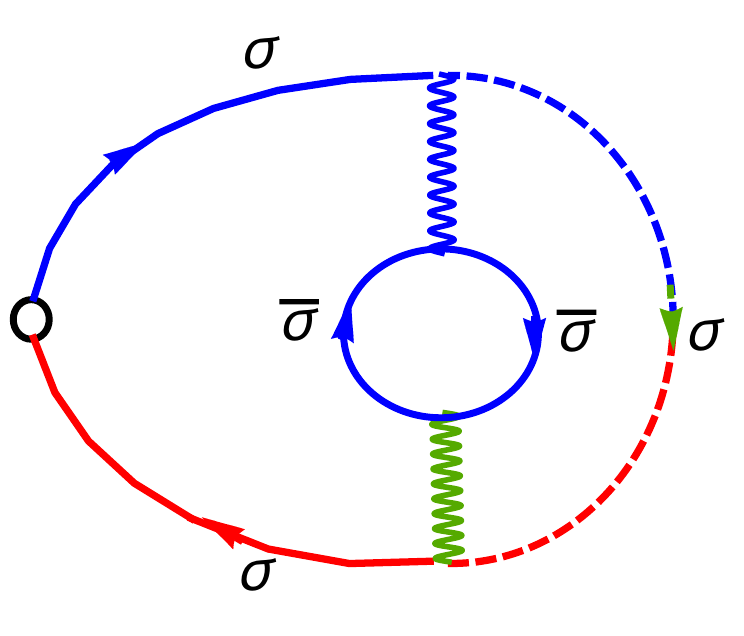}
\caption{(Color online) The $\phi_e\Phi$ type-2 current correction diagram (Left panel) and the corresponding diagram with the splitting of local $\Phi$ vertices (Right panel). In the diagram the $\Phi$ vertex consist in the Pauli matrices product $\tau_{\sigma\sigma}.\tau_{\bar{\sigma}\bar{\sigma}}{=}-1$. The diagram contain the even number of fermionic loops (two) and hence no extra negative sing occur due to the fermionic loop.  The $\Phi$ vertex has the renormalization factor of $-\frac{1}{2}$. Hence the overall weight factor of this diagram is $-\frac{1}{2}{\times} (-1)$ as will be seen in Fig.~\ref{LRFIG} and Fig.~\ref{BLRFIG}.}\label{weightfactor1}
\end{figure}
\begin{figure}[t]
\includegraphics[width=2.8cm, valign=c]{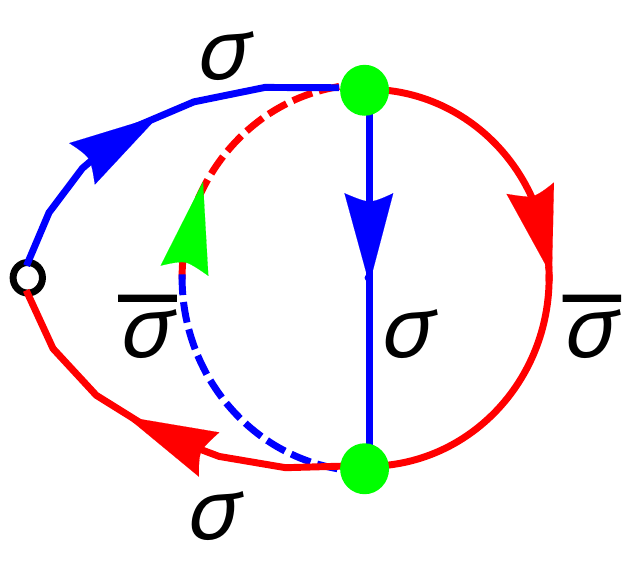}\;\;
\includegraphics[width=2.8cm, valign=c]{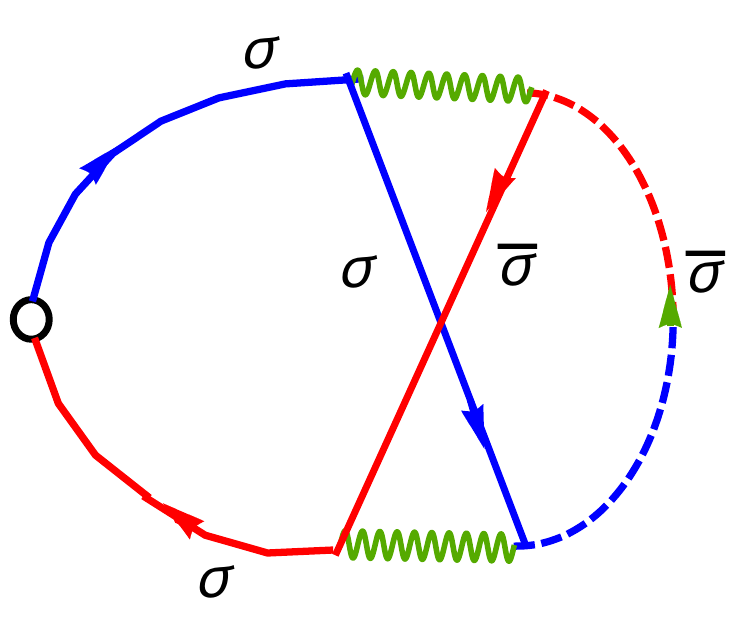}
\caption{(Color online) The $\Phi^2$ type-2 current correction diagram (Left panel) and the corresponding diagram with the splitting of local $\Phi$ vertices (Right panel). In the diagram the upper $\Phi$ vertex consist in the Pauli matrices product 
$\tau_{\sigma\sigma}.\tau_{\bar{\sigma}\bar{\sigma}}{=}-1$. Similarly the lower $\Phi$ vertex contain the product of $\tau_{\bar{\sigma}\sigma}.\tau_{\sigma\bar{\sigma}}{=}2$. The diagram contain no fermionic loops and hence no extra negative sing occur due to the fermionic loop. Each $\Phi$ vertex has the renormalization factor of $-\frac{1}{2}$. Hence the overall weight factor of this diagram is $\frac{1}{4}{\times} (-2)$ as will be seen in Fig.~\ref{LRFIG} and Fig.~\ref{BLRFIG}}\label{weightfactor2}
\end{figure}
However, in these equations proper weight factors which emerge from (i) the number of closed fermionic loops, (ii) SU(2) algebra of Pauli matrices and (iii) additional factors originating from the definition of the FL constants in the Hamiltonian (the extra factor of $-1/2$ in $H_{\Phi}$) are still missing and are accounted for separately. 
As a result our final expression for the second-order perturbative interaction corrections to the current is given by (see Appendix~\ref{NC})
\begin{align}\label{currentin}
\frac{\delta I_{\text{in}}}{2e^2V/h}&=\left[\frac{2}{3}(\phi^2_e+\phi^2_o)+3\Phi^2-2(\phi_e+\phi_o)\Phi\right](\pi T)^2\nonumber\\
&+\Big[\frac{5}{12}(\phi^2_e+\phi^2_o)+{\color{black}3}\Phi^2-{\color{black}2}(\phi_e+\phi_o)\Phi\nonumber\\
&{\color{black}+}\frac{1}{2}\phi_e\phi_o\Big](e V)^2.
\end{align}
\noindent The first term $\propto (\pi T)^2$ in Eq.~\eqref{currentin} is the linear response 
result given by type-1 and type-2 diagrams. The second term (surviving also at $T=0$)
is the non-linear response contribution arising from all type 1-4 diagrams. The inelastic current Eq.~\eqref{currentin} vanishes at the symmetry point. Moreover the linear response and the non-linear response contributions vanish at the symmetry point independently. Also the elastic and inelastic currents approach zero separately when the system is fine-tuned
to the symmetry point. These properties will be reproduced in arbitrary order of perturbation theory.
\section{Transport properties}\label{tp}
\noindent The total current consists of the sum of elastic and inelastic parts which upon using the FL-identity $\alpha_a{=}\phi_a$ takes the form
\begin{align}\label{totalcurr}
\frac{\delta I}{2e^2V/h}&=\left[(\pi T)^2 {\color{black}+}(eV)^2\right]3(\Phi-\frac{2}{3}\alpha_e)(\Phi-\frac{2}{3}\alpha_o)\nonumber\\
&+\left[B^2 +(\pi T)^2 +\frac{1}{2}(eV)^2\right]\left(\alpha_e-\alpha_o\right)^2 .
\end{align}
This Eq.~\eqref{totalcurr} constitutes the main result of this work where the second term describes universal behaviour \cite{GP_Review_2005} scaled with $\left(1/T_K^e{-}1/T_K^o\right)^2$, 
while the first one, containing an extra dependence on the ratio $T_K^o/T_K^e$ accounts for the non-universality associated with the lack of conformal symmetry away from the symmetry-protected points. The Eq.~\eqref{totalcurr} demonstrates the magnetic field $B$, temperature $T$ and voltage $V$ 
behaviour of the charge current characteristic for the Fermi-liquid systems.
Therefore, following \cite{HWDK_PRB_(89)_2014} we introduce general FL constants as follows:{\color{black}
\begin{equation}\label{cbtv}
\frac{1}{G_0}\frac{\partial I}{\partial V}=c_B B^2 + c_T (\pi T)^2 +c_V (eV)^2.
\end{equation}
}
\begin{align}
\frac{c_T}{c_B}=1+3\mathcal{F},\;\;\;\;\;\;
\frac{c_V}{c_B}=\frac{{\color{black} 3}}{2}+{\color{black}9}\mathcal{F}.
\end{align}
\noindent Here the parameter 
\begin{align}
\label{eq:define-F}
\mathcal{F}=\frac{(\Phi-\frac{2}{3}\alpha_e)(\Phi-\frac{2}{3}\alpha_o)}{\left(\alpha_e-\alpha_o\right)^2 }=\frac{4}{9}\frac{(\lambda_{eo}-\lambda_e)(\lambda_{eo}-\lambda_o)}{\left(\lambda_e-\lambda_o\right)^2 }.
\end{align}
{\color{black} The parameter $\mathcal{F}$} vanishes in the limit of strong asymmetry, 
${\color{black}\lambda_{eo}{\ll}\lambda_e{\ll}\lambda_o}$
 in which the ratios 
\begin{align}
\label{eq:FermiLiquidRatiosAsymmetry}
\left. c_{T}/{c_B}\right|_{\color{black}\lambda_{eo}\ll\lambda_e\ll\lambda_o}{=}1,\;\;\;\;\;\;
\left. c_{V}/{c_B}\right|_{\color{black}\lambda_{eo}\ll\lambda_e\ll\lambda_o}{=} 3/2
\end{align} 
correspond to the universality class of the single-channel Kondo model
\cite{Glazman_PRL_2001,GP_Review_2005}.

On the other hand, near the symmetry point
$\lambda_e {=} \lambda_o {=} \lambda_{eo}$, 
the function $\mathcal{F}$ evidently depends sensitively on the precise
manner in which the symmetry point is approached. In fact, \textit{a
  priori} it appears unclear whether $\mathcal{F}$ even reaches a
well-defined value at this point. To clarify this, additional
information on the parameters $\lambda_e$, $\lambda_o$ and
$\lambda_{eo}$ is} {\color{black} required.}

\begin{figure}[t]
\begin{center}
 \includegraphics[width=60mm]{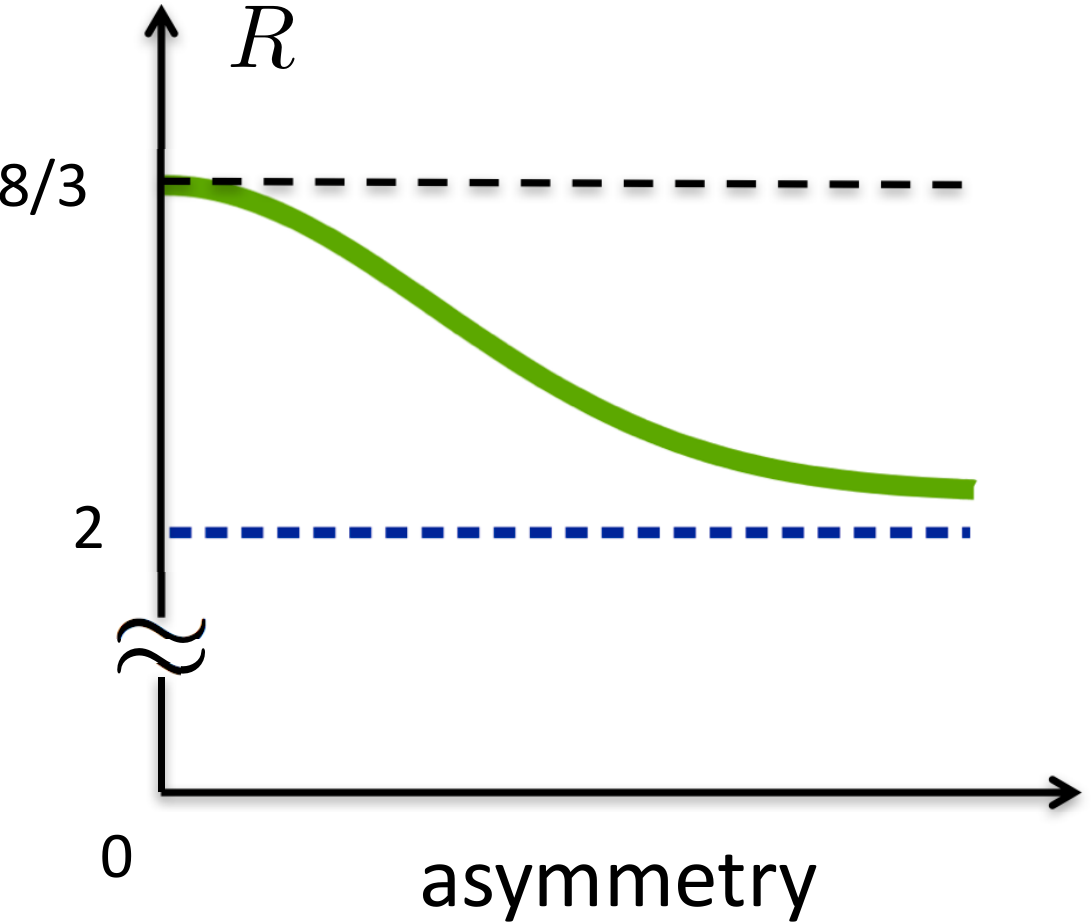}
 \caption{(Color online) Cartoon sketching the evolution of the Wilson
   ratio as a function of 
\jvd{increasing ``asymmetry'',
     meaning that the ratios 
      $\lambda_e/\lambda_o$ and
     $\lambda_{eo}/\lambda_e$ both decrease from 1 at the left to
     0 at the right.}  When \jvd{$\lambda_e{=}\lambda_o{=}\lambda_{eo}$,
     meaning that} the even and odd Kondo temperatures coincide, the
   total spin current is conserved \cite{Affleck_Lud_PRB(48)_1993}
   {\color{black} and $R{=}8/3$ \cite{AFFLECK_AP_1995}.}
   In the limit of extremely
   (exponentially) strong channel asymmetry of 2SK model, the (C)
   regime shown on Fig. \ref{cart.2} shrinks to zero. As a result, the
   1CK universality class appears and {\color{black} Wilson Ratio is
   $R{=}2$ \cite{AFFLECK_AP_1995}}. The behaviour of the Wilson ratio between these to limits is
   \jvd{presumably} monotonic, since the 2SK model has no other strong
   coupling fixed points. }
\label{WilsonRatio}
\end{center} 
\end{figure}

\cm{In full generality, the  three parameters  $\lambda_e$, $\lambda_o$ and $\lambda_{eo}$ of the FL theory are independent from each other. Nonetheless, we are considering here a specific Hamiltonian Eq.~\eqref{Heff} with only two independent parameters $J_e$ and $J_o$, which implies that $\lambda_{eo}$ is in fact a function of $\lambda_e$ and $\lambda_o$. Although the corresponding functional form is not known, it can be deduced in the vicinity of the symmetric point $\lambda_e {=} \lambda_o {=} \lambda_{eo}$ from the following argument: the obvious $e\leftrightarrow o$ symmetry imposes that the Wilson ratio $R{=}8/3$ is an extremum at the symmetric point (see Fig. \ref{WilsonRatio}), or else said, that its derivative with respect to the channel imbalance ratio $\lambda_o/\lambda_e$ vanishes. The only expression compatible with this requirement and the $e\leftrightarrow o$ symmetry is $\lambda_{eo}{=}(\lambda_e {+} \lambda_o)/2$, valid in the immediate vicinity of the symmetry point. Inserting this dependence in Eq.~\eqref{eq:define-F} predicts $\lim_{\lambda_e\to \lambda_o}\mathcal{F} {=} - 1/9$ at the symmetric point, and}

\begin{align}
\label{eq:FermiLiquidRatiosSymmetry}
\left. c_{T}/{c_B}\right|_{\color{black}\lambda_{eo}=\lambda_e=\lambda_o}{=} 2/3,\;\;\;\;\;\;
\left. c_{V}/{c_B}\right|_{\color{black}\lambda_{eo}=\lambda_e=\lambda_o}{=} {\color{black}1/2}.
\end{align}
{\color{black} To summarize, under the assumption that the Wilson 
ratio is maximal at the symmetry point, we have arrived
at the following conclusion: 
as the degree of asymmetry is reduced,
i.e.\ the ratios ${\color{black} \lambda_e/\lambda_o}$ and ${\color{black} \lambda_{eo}/\lambda_e}$ 
increased from $0$ to $1$, 
the ratios of Fermi liquid coefficients 
$ c_{T}/{c_B}$ and $ c_{V}/{c_B}$ {\color{black} decrease} from 
the {\color{black} maximal} values of Eq.~\eqref{eq:FermiLiquidRatiosAsymmetry},
to the {\color{black} minimal} values of Eq.~\eqref{eq:FermiLiquidRatiosSymmetry},
characteristic of the 1CK and 2SK fixed points, respectively.}
 
\section{Discussion}\label{diss}
We constructed a Fermi-liquid theory of a two-channel,
two-stage Kondo model when both scattering channels are close to the
resonance. This theory completely describes the transport in in- and
out-of-equilibrium situation of the 2SK model. The elastic and
inelastic contributions to the charge current through the 2SK model
have been calculated using the full-fledged non-equilibrium Keldysh
formalism for arbitrary relation between two Kondo energy
scales. While computing the current correction, we performed the full
classification of the Feynman diagrams for the many-body perturbation
theory on the Keldysh contour. We demonstrated the cancellation of the
charge current at the symmetry protected point. The linear response
and beyond linear response contributions to the current vanish
separately at the symmetry point. Moreover, the independent
cancellation of the elastic and inelastic currents at the symmetry
protected point was verified. The theoretical method developed in the paper
provides a tool for both quantitative and qualitative description of
charge transport in the framework of the two-stage Kondo problem.  In
particular, the two ratios of FL constants, $c_T/c_B$ and $c_V/c_B$,
quantify the ``amount" of interaction between two channels. The
interaction is strongest at the symmetry protected point due to strong
coupling of the channels. The interaction is weakest at single-channel
Kondo limit where the odd channel is completely decoupled from the even
channel. While we illustrated the general theory of two resonance
scattering channels by the two-stage Kondo problem, the formalism
discussed in the paper is applicable for a broad class of models
describing quantum transport through nano-structures \cite{qpc_nat,
  qpc_jvd, Meir_NAT_(442)_2006} and behaviour of strongly correlated
systems \cite{Coleman_book}.

As an outlook, the approach presented in this paper can be applied to
the calculation of current-current correlation functions (charge
noise) of the 2SK problem and, by computing higher cumulants of the
current, to studying the full-counting statistics \cite{Levitov_JETP,
Levitov}. It is straightforward to extend the presented ideas for
generic Anderson-type models away from the particle-hole symmetric
point \cite{Oguri_Hewson_2017(1),Oguri_Hewson_2018(2),Oguri_Hewson_2018(3)}, and generalize it for the SU(N) Kondo impurity
\cite{Karki_MK_TE} and multi-terminal (multi-stage) as well as
multi-dot setup. The general method developed in the paper is not
limited by its application to charge transport through quantum
impurity --- it can be equally applied to detailed description of the
thermo-electric phenomena on the nano-scale \cite{Karki_MK_TE}.
\section*{Acknowledgements}
We thank Ian Affleck, Igor Aleiner, Boris Altshuler,  Natan Andrei, Andrey Chubukov, Piers Coleman, Leonid Glazman, Karsten Flensberg, Dmitry Maslov, Konstantin Matveev, Yigal Meir, Alexander Nersesyan, Yuval Oreg, Nikolay Prokof'ev and Subir Sachdev for fruitful discussions. We are grateful to Seung-Sup Lee for discussions and sharing his preliminary results on a numerical study of multi-level Anderson and Kondo impurity models.
This work was finalized at the Aspen Center for Physics, which was supported by National Science Foundation Grant No.PHY-1607611 and was partially supported (M.N.K.) by a grant from the Simons Foundation.
J.v.D. was supported by the Nanosystems Initiative Munich.
D.B.K and M.N.K appreciate the hospitality of the Physics Department, Arnold Sommerfeld Center for Theoretical Physics and Center for NanoScience, Ludwig-Maximilians-Universit{\"a}t M{\"u}nchen, where part of this work has been performed.
\appendix

\section{Overview of flow from weak to strong coupling}
\subsection{Weak coupling regime}\label{weak_cop_reg}
We assume that at sufficiently high temperatures (a precise definition of this condition is given below) the even and odd channels 
do not talk to each other. As a consequence, we renormalise the coupling between channels and impurity spins ignoring the cross-channel interaction.
Performing Anderson's poor man's scaling procedure \cite{Anderson} to the even and odd channels independently we obtain the system of two decoupled renormalization group (RG) equations:
\begin{equation}\label{RG}
\frac{dJ_e}{d\Lambda}=2N_FJ^2_e, \quad \frac{dJ_o}{d\Lambda}=2N_FJ^2_o,
\end{equation}
where $N_F$ is the 3D-density of states in the leads.
The parameter $\Lambda{=}\ln\left(\frac{D}{\varepsilon}\right)$ depends on
the ultraviolet cutoff of the problem (conduction bandwidth $D$).
Note that the RG Eqs.~\eqref{RG} are decoupled only in one-loop approximation
(equivalent to a summation of so-called parquet diagrams). The solution of these RG equations defines two characteristic energy scales, namely 
$T_K^a {=} D\exp\left(-1/(2N_F J_a)\right)$, which are the Kondo temperatures in the even and odd channels respectively. The second loop corrections to RG couple the equations, {\color{black} generating the cross-term  
$\propto -J_{eo}\;\mathbf{s}_e\cdot\mathbf{s}_o$ with 
$J_{eo}\sim N_F J_e\cdot J_o$.
This emergent term flows under RG and becomes one of the leading irrelevant operators of the strong coupling fixed point
(the others are $:\mathbf{s}_e\cdot\mathbf{s}_e:$ and $:\mathbf{s}_o\cdot\mathbf{s}_o:$, see Eq.\ref{lir}).}
 In addition, the second-loop corrections to RG lead to a renormalization of the pre-exponential factor in the definition of the Kondo temperatures.

Summarizing, we see that the $S{=}1$, ${\cal K}{=}2$ fully screened Kondo model
has a unique strong coupling fixed point, where couplings $J_e$ and $J_o$ diverge in the RG flow.
This strong coupling fixed point falls into the FL universality class.
The  weak coupling regime is therefore defined as 
$(B, T, eV){\gg} (T_K^e, T_K^o)$. Since the interaction between the even channel and local impurity spin corresponds to the maximal eigenvalue of the matrix
Eq.~\eqref{exmat}, we will assume below that the condition $T_K^e {\geqslant} T_K^o$ holds for any given $B, T$ and $eV$ and, we thus define $T^{\text{min}}_K{=}T^o_K$. The differential conductance decreases monotonically with increasing temperature in the weak-coupling regime (see Fig.~\ref{cart.2}) being fully described by the perturbation theory \cite{GP_Review_2005}
in $[1/\ln(T/T_K^e), 1/\ln(T/T_K^e)] \ll 1$.
\subsection{Intermediate coupling regime}\label{int_cop_reg}
Next we consider the intermediate coupling regime $T_K^o {\leqslant} (B, T, eV) {\leqslant} T_K^e$ depicted as the {\color{black} characteristic hump} in Fig.~\ref{cart.2}. Since the solution of one-loop RG 
Eqs.~\eqref{RG} is given with logarithmic accuracy, we assume without loss of generality that $T_K^e$ and $T_K^o$ are of the same order of magnitude
unless a very strong (exponential) channel asymmetry is considered.
Therefore, the {\color{black} ``hump regime''} is typically very small
and the {\color{black} hump} does not have enough room to be formed. The intermediate regime
is characterized by an incomplete screening (see Fig.~\ref{cart.2})
when one conduction channels (even) falls into a strong coupling regime
while the other channel (odd) still remains at the weak coupling.
Then the strong-coupling Hamiltonian for the even channel is derived along the lines of Affleck-Ludwig paper Ref. \cite{Affleck_Lud_PRB(48)_1993} and is given by:
\begin{align}
H_{\rm even}&{=}H^e_0{+}\frac{3}{2}\lambda_e \rho_{e\uparrow}\rho_{e\downarrow}{-}\frac{3}{4v_F}\lambda_e \sum_{kk'\sigma} \left(\varepsilon_k+\varepsilon_k'\right) b^{\dagger}_{ek\sigma}b_{ek'\sigma},
\end{align} 
where the $b$-operators describe Fermi-liquid excitations, $\rho_{e\sigma}(x{=}0){=}\sum_{kk'}b^\dagger_{ek \sigma}b^\pdag_{ek' \sigma}$
and $\lambda_e\propto 1/T_K^e$ is the leading irrelevant coupling constant \cite{Affleck_Lud_PRB(48)_1993}.

The weak-coupling part of the remaining Hamiltonian is described by a $s_{\rm imp}{=}1/2$ Kondo-impurity Hamiltonian $H_{\rm odd}{=}J_o{\textbf s}_o\cdot {\textbf s}_{\rm imp}$. Here we have already taken into account that the impurity spin is partially screened by the even channel during the first stage process of the Kondo effect. We remind that the coupling between the even and odd channels is facilitated by a {\it ferromagnetic} interaction which emerges, {\color{black} being however irrelevant} in the intermediate coupling regime. 
Thus, the differential conductance
does reach a maximum $G/G_0{\approx} 1$ {\color{black} with a characteristic hump} \cite{Glazman_PRL_2001}, \cite{Coleman_PRB(75)_2007} at the intermediate coupling regime. 
Corresponding corrections 
(deviation of the conductance at the top of the {\color{black} hump} from the unitary limit
$G_0{=}2e^2/h$) can be calculated with logarithmic accuracy 
$|\delta G/G_0|{\propto} 1/\ln^2(T_K^e/T_K^o)$
\cite{Nozieres_Blandin_JPhys_1980}, \cite{Anderson} (see also review \cite{GP_Review_2005} and \cite{Coleman_PRB(75)_2007}
for details).
\section{Counterterms}\label{counter}
We proceed with the calculation of the corrections to the current by eliminating the dependence on the cutoff parameter $D$ by adding the counter terms in the Hamiltonian \cite{Affleck_Lud_PRB(48)_1993, Mora_Clerk_Hur_PRB(80)_2009}
\begin{equation}\label{ct}
H_{c}=-\frac{1}{2\pi\nu}\sum_a\sum_{kk'\sigma} \left(\delta \alpha_a+\delta \Phi\right) (\varepsilon_k+\varepsilon_{k'}):b^{\dagger}_{ak\sigma}b_{ak'\sigma}:,
\end{equation}
so that we consider only the contribution which remain finite for 
$D{\rightarrow}\infty$.
The Eq.~\eqref{ct} corresponds to the renormalization of leading irrelevant coupling constant $\alpha_a$ such that $\alpha_a\rightarrow \alpha_a+\delta \alpha_a+\delta \Phi$ with
\begin{align}
\delta \alpha_a=&-\alpha_a\phi_a \frac{6D}{\pi}\log\left(\frac{4}{3}\right).\\
\delta \Phi=&-\Phi^2\frac{9D}{\pi}\log\left(\frac{4}{3}\right).
\end{align}
During the calculation of the interaction correction we neglected those terms which produce the contribution proportional to the cutoff $D$ [for example, $\propto \int \frac{d\varepsilon}{2\pi}\left(\Sigma^{++}(\varepsilon)-\Sigma^{--}(\varepsilon)\right)i\pi\nu\Delta f(\varepsilon)$]. This renormalization of leading irrelevant coupling constant Eq.~\eqref{ct} exactly cancel these terms. 
\section{Elastic current}\label{ec}
To get the elastic current Eq.~\eqref{elastic}, we start from the Landauer-B\"uttiker formula Eq.~\eqref{buttiker}
\begin{equation}\label{buttiker1}
 I_{\text{el}}=\frac{2 e}{h} \int_{-\infty}^{\infty} d\varepsilon T(\varepsilon)\Delta f(\varepsilon),
\end{equation}
where the energy dependent transmission coefficient, $T(\varepsilon){=}\frac{1}{2}\sum_\sigma\sin^2(\delta^e_\sigma(\varepsilon)-\delta^o_\sigma(\varepsilon))$ and $\Delta f(\varepsilon){=}f_L(\varepsilon)-f_R(\varepsilon)$.
Taylor expanding the phase shifts to the first order in energy and retaining only upto second order in energy terms in the $T(\varepsilon)$, we arrive at the expression

\begin{equation}\label{cu1}
I_{\text{el}}=\frac{2 e}{h}  (\alpha_e-\alpha_o)^2\int_{-\infty}^{\infty} d\varepsilon \varepsilon^2 \Delta f(\varepsilon).
\end{equation}
\noindent  To compute the integral Eq.~\eqref{cu1} we use the property of the Fourier transform. For the given function $\Delta f(\varepsilon)$, it's Fourier transform is defined as
\begin{equation}\label{cu3}
\Delta f(t)=\frac{1}{2\pi} \int_{-\infty}^{\infty}  e^{-i\varepsilon t}\Delta f(\varepsilon) d\varepsilon.
\end{equation}
Taking $n$-th derivative of Eq.~\eqref{cu3} at $t=0$ we get
\begin{equation}\label{ft}
\int_{-\infty}^{\infty}  \varepsilon^n  \Delta f(\varepsilon) d\varepsilon= \frac{2\pi}{(-i)^n}
\left. \partial_t^n \left[\Delta f(t)\right]\right|_{_{t=0}}.
\end{equation}
\noindent Substituting Eq.~\eqref{ft} for $n=2$ into Eq.~\eqref{cu1}, the elastic current cast into the form
\begin{equation}\label{cu2}
I_{\text{el}}=\frac{2 e}{h}  (\alpha_e-\alpha_o)^2(-2\pi)\left. \partial_t^2 \left[\Delta f(t)\right]\right|_{_{t=0}}.
\end{equation}
\noindent The Fourier transform of $\Delta f(\varepsilon)$ for $\mu_{L/R}=\pm eV/2$ is defined by
\begin{equation}\label{delf}
\Delta f(t)=T \frac{\sin(\frac{eVt}{2})}{\sinh(\pi T t)}.
\end{equation}
Using Eq.~\eqref{delf} into Eq.\eqref{cu2}, we can easily arrive at the expression Eq.~\eqref{elastic} for the elastic current
at finite temperature $T$, finite bias voltage $V$ and finite in-plane (Zeeman) magnetic field $B$ (assuming $(T,eV,B)\ll T_K^o$)
\begin{equation}\label{elastic1}
\frac{I_{\text{el}}}{2e^2V/h}=\left[B^2 + \frac{(eV)^2}{12}+ \frac{(\pi T)^2}{3}\right] (\alpha_e-\alpha_o)^2.
\end{equation}

\section{Net electric current}\label{NC}
Here we present the detail of the computation of total electric current (sum of elastic and inelastic parts) given by Eq.~\eqref{totalcurr}. We discuss the total current in linear-response (LR) and beyond linear-response (BLR) regime separately. The elastic part is given by Eq.~\eqref{elastic} and the inelastic part which is composed of the four types of diagrams is expressed by Eq.~\eqref{currentin}.
\subsection{Linear Response (LR)}
\begin{figure}[h]
\begin{equation}\nonumber
\begin{split}
\delta I^{\phi^2}_{\rm in}&=\phantom{-}\phantom{-}\left[\underbrace{
\includegraphics[width=1.8cm, valign=c]{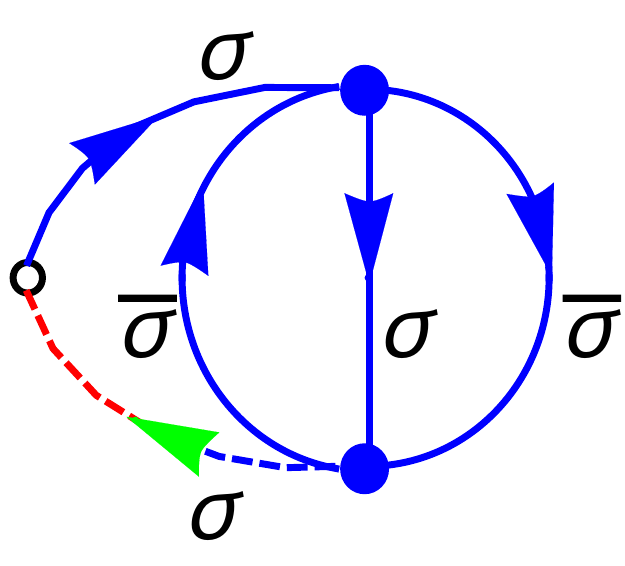}
}_{\rm type-1}\right]+\delta I^{\phi^2_o}_{\rm in}\\
&=\frac{2e^2V}{h}\left[A^{(1)}_V(eV)^2+A^{(1)}_T(\pi T)^2\right]\left(\phi^2_e+\phi^2_o\right)\\
\delta I^{\Phi^2}_{\rm in}&=+\frac{1}{4}\left[\underbrace{\phantom{-}
\includegraphics[width=1.6cm, valign=c]{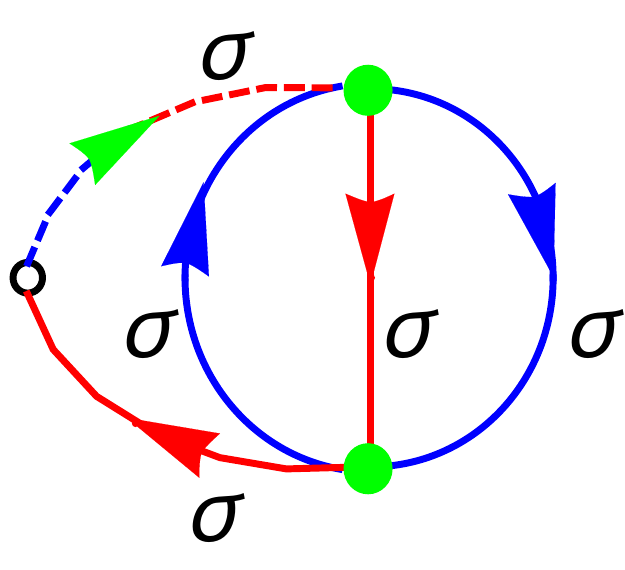}
+\includegraphics[width=1.6cm, valign=c]{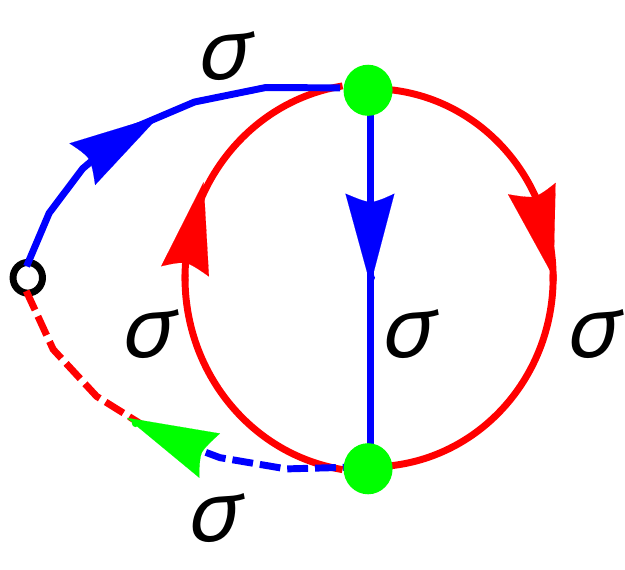}
+\includegraphics[width=1.6cm, valign=c]{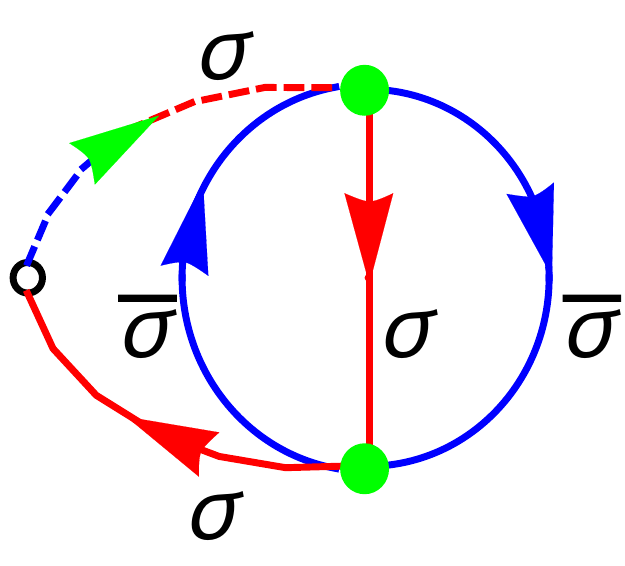}}_{\rm \text{type-1}}
\right]\\
&\phantom{-}+\frac{1}{4}\left[\underbrace{\includegraphics[width=1.6cm, valign=c]{y5.pdf}
+4\includegraphics[width=1.6cm, valign=c]{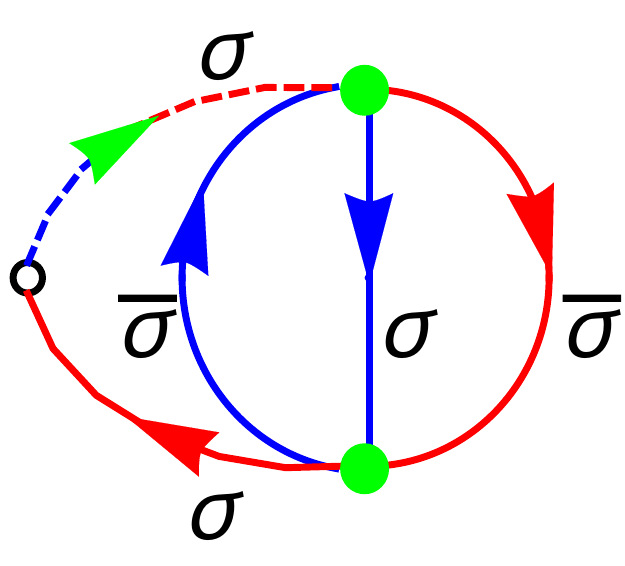}
+4\includegraphics[width=1.6cm, valign=c]{y7.pdf}}_{\rm \text{type-1}}\right]\\
&\phantom{-}+\frac{1}{4}\left[\underbrace{
\includegraphics[width=1.6cm, valign=c]{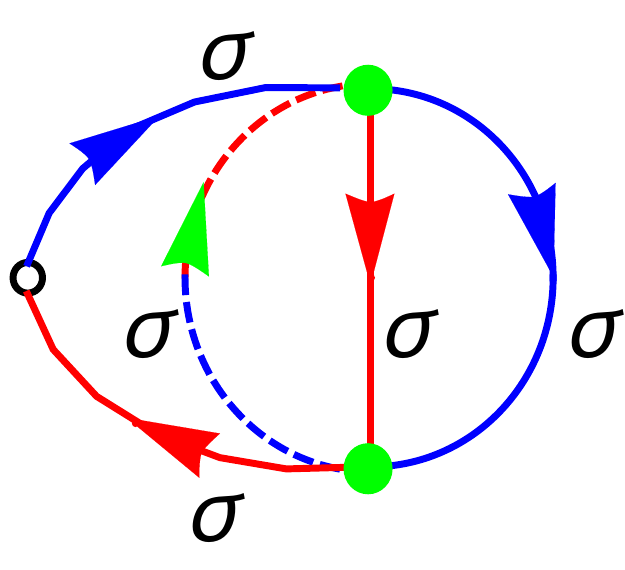}
-2\includegraphics[width=1.6cm, valign=c]{y9.pdf}
-2\includegraphics[width=1.6cm, valign=c]{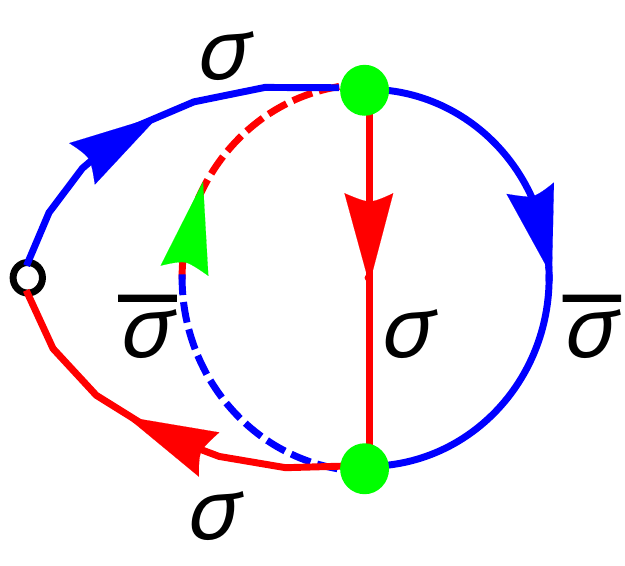}}_{\rm \text{type-2}}\right]\\
&=\frac{6e^2V}{h}\Big[({A^{(1)}_V}{-}\frac{A^{(2)}_V}{4}){(}eV{)}^2{+}(A^{(1)}_T{-}\frac{A^{(2)}_T}{4})(\pi T)^2\Big]\Phi^2\\
\delta I^{\phi_a\Phi}_{\rm in}&={-}\frac{1}{2}\left[\underbrace{
-\includegraphics[width=1.7cm, valign=c]{y11.pdf}
-2\includegraphics[width=1.7cm, valign=c]{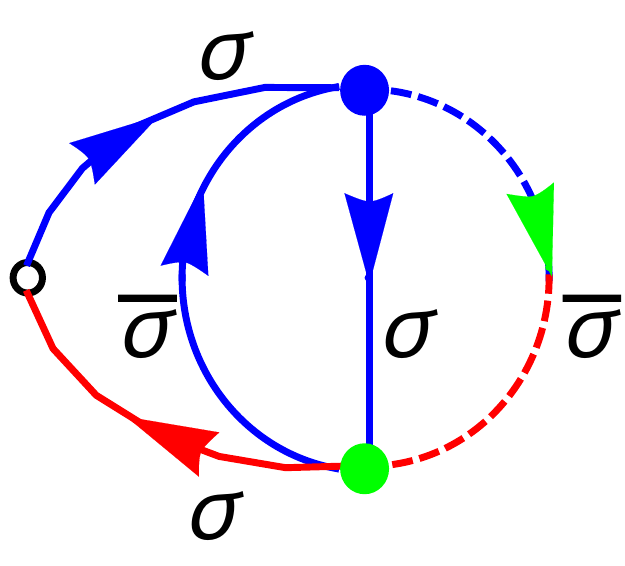}}_{\rm \text{type-2 }}
\right]+\delta I^{\phi_o\Phi}_{\rm in}\\
&=\frac{3e^2V}{h}\left[A^{(2)}_V(eV)^2+A^{(2)}_T(\pi T)^2\right]\left(\phi_e+\phi_o\right)\Phi
\end{split}
\end{equation}
\caption{(Color online) Feynman diagrams of type-1 and type-2 contributing 
to the charge current both in the linear response and beyond the linear response regime. The coefficients computed in the Sec \ref{ccB1} and Sec \ref{ccB2} take the following values: $A^{(1)}_T{=}2/3$, $A^{(2)}_T{=}-4/3$,
$A^{(1)}_V{=}5/12$, $A^{(2)}_V{=}-5/6$.}\label{LRFIG}
\end{figure}

As discussed in the main text, both elastic and inelastic processes contribute to the LR current. The LR contribution of the elastic part is expressed by Eq.~\eqref{elastic}. The diagrams of type-1 and type-2 has the finite linear response contribution to the inelastic current. As detailed in Fig.~\ref{LRFIG}, we have the expression of total linear response current
\begin{align}\label{current2nd}
&\frac{\delta I^{\text{LR}}}{2e^2V/h}\frac{1}{(\pi T)^2}=\Big[\underbrace{\frac{1}{3}\left(\alpha_e-\alpha_o\right)^2}_{\rm \text{LR elastic part}}\Big]+
\nonumber\\&\Big[\underbrace{A^{(1)}_T(\phi^2_e{+}\phi^2_o){+}3A^{(1)}_T\Phi^2
{+}\frac{3A^{(2)}_T}{2}(\phi_e{+}\phi_o)\Phi{-}\frac{3A^{(2)}_T}{4}\Phi^2}_{\rm \text{LR inelastic part (type-1 and type-2 diagrams)}}\Big]\nonumber\\
&{=}\left[\frac{1}{3}\left(\alpha_e{-}\alpha_o\right)^2{+}\frac{2}{3}(\phi^2_e{+}\phi^2_o){-}2(\phi_e{+}\phi_o)\Phi{+}3\Phi^2\right]\nonumber\\
&{=}\left[(\alpha_e-\alpha_o)^2 
+3(\Phi-\frac{2}{3}\alpha_e)(\Phi-\frac{2}{3}\alpha_o)\right].
\end{align}
At the symmetry point the linear response contribution to the current given by the Eq.~\eqref{current2nd} exactly vanishes.

\subsection{Beyond Linear Response (BLR)} 
\begin{figure}[h]
\begin{equation}\nonumber
\begin{split}
\delta I^{\Phi^2}_{\rm in}&{=}\phantom{-}
\frac{1}{4}\left[\underbrace{\phantom{-}\includegraphics[width=1.7cm, valign=c]{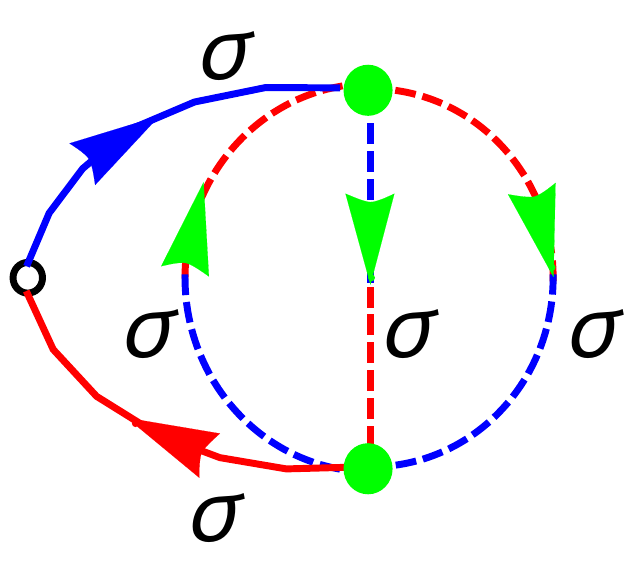} 
{+}\phantom{-}\includegraphics[width=1.7cm, valign=c]{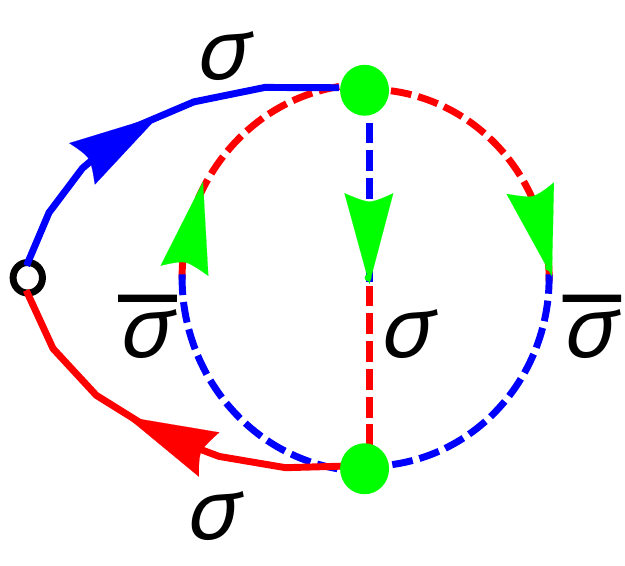}
  {+}4\includegraphics[width=1.7cm, valign=c]{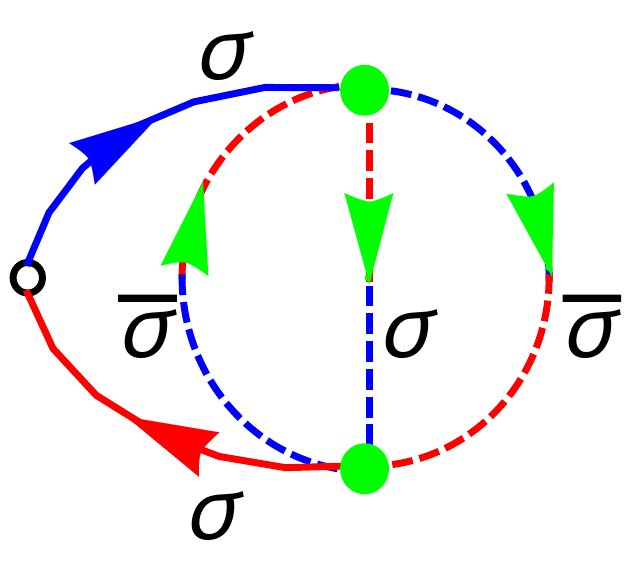}}_{\rm \text{type-4}}\right]\\
&\phantom{-}{+}\frac{1}{4}\left[\underbrace{\phantom{-}
 \includegraphics[width=1.7cm, valign=c]{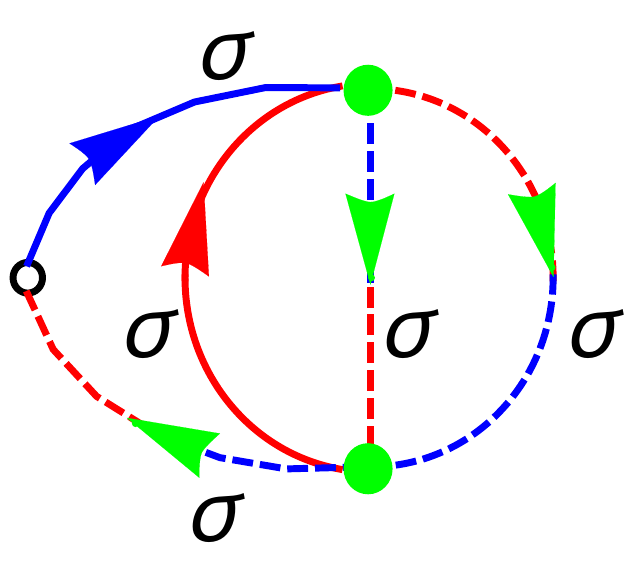}
 {+}\phantom{-}\includegraphics[width=1.7cm, valign=c]{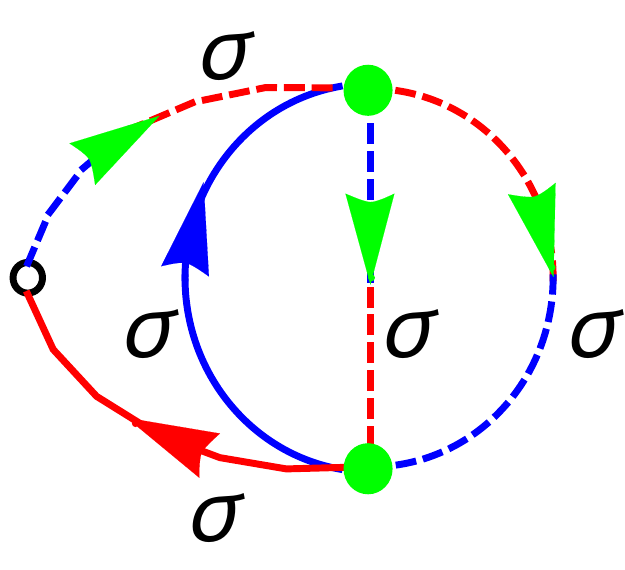}
 {-}2\includegraphics[width=1.7cm, valign=c]{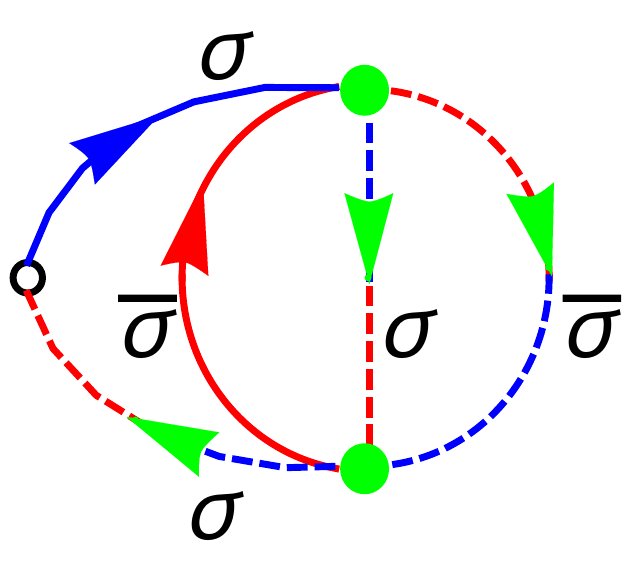}}_{\rm \text{type-3}}
\right]\\
&\phantom{-}{+}\frac{1}{4}\left[\underbrace{-2\includegraphics[width=1.7cm, valign=c]{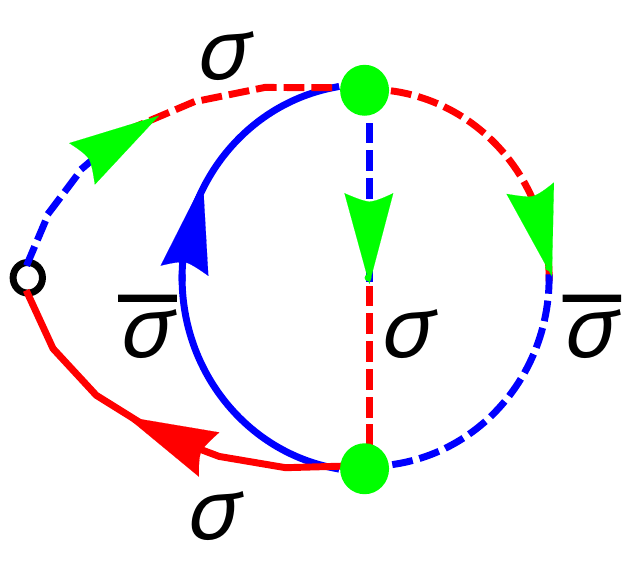}
 {-}2\includegraphics[width=1.7cm, valign=c]{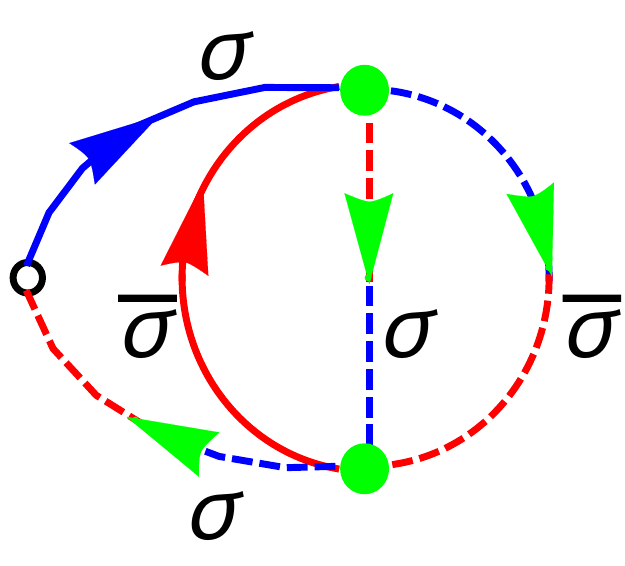}
 {-}2\includegraphics[width=1.7cm, valign=c]{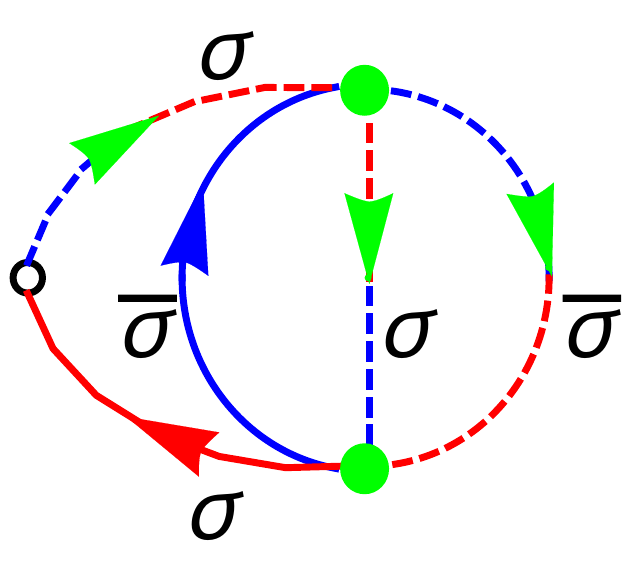}}_{\rm \text{type-3}}
\right]\\
&{=}\frac{3e^2V}{h}\left[(A^{(4)}_V{-}A^{(3)}_V)(eV)^2+(A^{(4)}_T{-}A^{(3)}_T)(\pi T)^2\right]{\Phi^2}\\
\delta I^{\phi_a\Phi}_{\rm in}&{=}{-}\frac{1}{2}\left[\underbrace{
 {-} \includegraphics[width=1.8cm, valign=c]{z10.pdf}
{-}\includegraphics[width=1.8cm, valign=c]{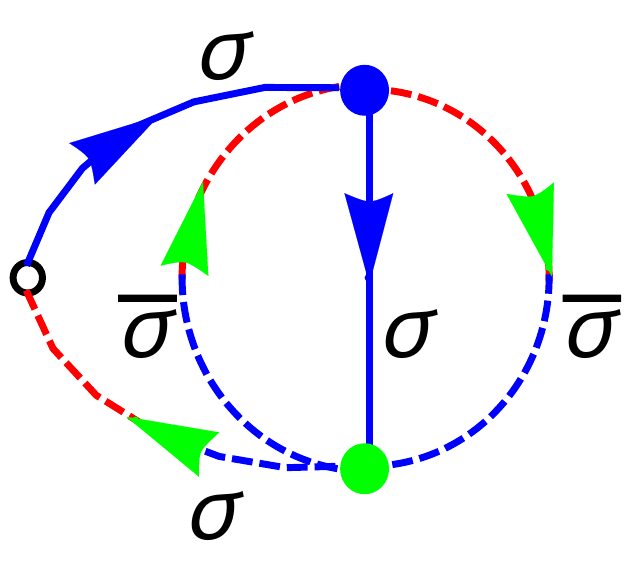}}_{\rm \text{type-3}}\right.\\
&\;\;\;\;\;\;\;\;\left.\underbrace{
{-}2\includegraphics[width=1.8cm, valign=c]{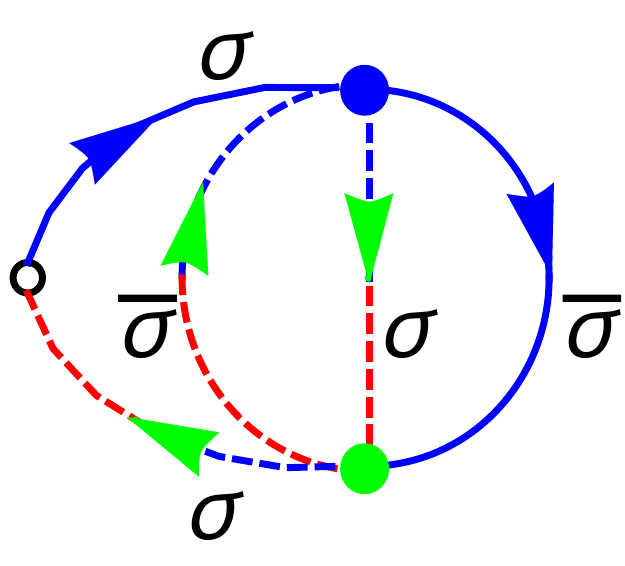}
{-}2\includegraphics[width=1.8cm, valign=c]{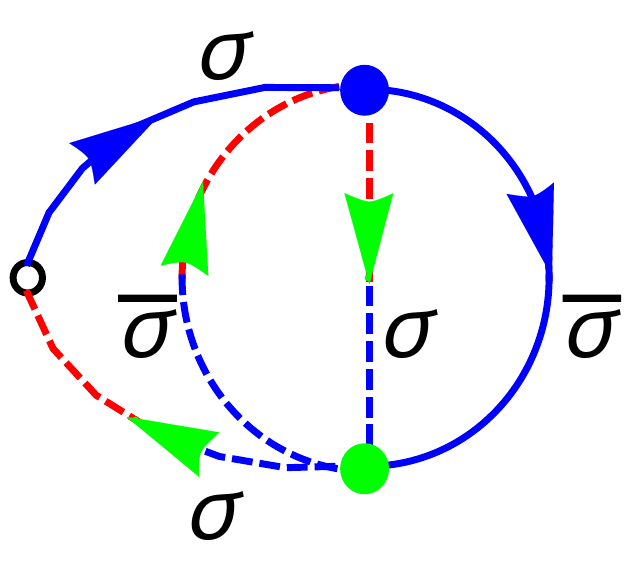}}_{\rm \text{type-3}}
\right]
+\delta I^{\phi_o\Phi}_{\rm in}\\
&{=}\frac{6e^2V}{h}\left[A^{(3)}_V(eV)^2{+}A^{(3)}_T(\pi T)^2\right]\left(\phi_e+\phi_o\right)\Phi\\
\delta I^{\phi_e\phi_o}_{\rm in}&{=}
\left[\underbrace{
\includegraphics[width=1.6cm, valign=c]{z14.pdf}}_{\rm type-4}
\right]{=}\frac{2e^2V}{h}\left[A^{(4)}_V(eV)^2{+}A^{(4)}_T(\pi T)^2\right]\phi_e\phi_o 
\end{split}
\end{equation}
\caption{(Color online) Feynman diagrams of type-3 and type-4 contributing 
to the charge current beyond the linear response. 
The coefficients computed in the Sec \ref{ccB3} and Sec \ref{ccB4} take the following values: $A^{(3)}_T{=}A^{(4)}_T{=}0$,
$A^{(3)}_V{=}{\color{black}-}1/4$, $A^{(4)}_V{=}{\color{black}1/2}$.
}\label{BLRFIG}
\end{figure}
The BLR contribution of the elastic part is expressed by Eq.~\eqref{elastic}. The diagrams of type-3 and type-4 produce the finite contribution to the inelastic current only beyond the LR regime. In addition to the LR contribution, the type-1 and type-2 diagrams also contribute to non-linear response. As detailed in Fig.~\ref{BLRFIG}, the total non-linear current is 
\begin{align}\label{current2nd1}
&\frac{\delta I^{\text{BLR}}}{2e^2V/h}\frac{1}{(eV)^2}=\Big[\underbrace{\frac{1}{12}\left(\alpha_e-\alpha_o\right)^2}_{\rm \text{BLR elastic part}}\Big]\nonumber\\
&{+}{\left[\underbrace{A^{(1)}_V(\phi^2_e{+}\phi^2_o){+}3A^{(1)}_V\Phi^2{+}\frac{3A^{(2)}_V}{2}(\phi_e{+}\phi_o)\Phi{-}\frac{3A^{(2)}_V}{4}\Phi^2}_{\rm \text{BLR inelastic part (type-1 and type-2 diagrams)}}\right]}\nonumber\\
&{+}{\left[\underbrace{A^{(4)}_V\phi_e\phi_o+3A^{(3)}_V\left(\phi_e+\phi_o\right)\Phi+\frac{3}{2}(A^{(4)}_V-A^{(3)}_V)\Phi^2}_{\rm \text{BLR inelastic part (type-3 and type-4 diagrams)}}\right]}\nonumber\\
&{=}
{\Big[\frac{1}{12}\left(\alpha_e-\alpha_o\right)^2+\frac{5}{12}(\phi^2_e+\phi^2_o)-\frac{5}{4}(\phi_e+\phi_o)\Phi}\nonumber\\
&\phantom{-}\phantom{-}{+\frac{15}{8}\Phi^2{\color{black}+}\frac{1}{2}\phi_e\phi_o{\color{black}-}\frac{3}{4}\left(\phi_e+\phi_o\right)\Phi{\color{black}+}\frac{9}{8}\Phi^2}\Big]\nonumber\\
&={\color{black}\left[\frac{1}{2}(\alpha_e-\alpha_o)^2+3(\Phi-\frac{2}{3}\alpha_e)(\Phi-\frac{2}{3}\alpha_o)\right]. }
\end{align}
The BLR contribution to the current expressed by Eq.~\eqref{current2nd1} goes to zero at the symmetry point $\alpha_e{=}\alpha_o{=}3\Phi/2$. 

The sum of the LR and BLR contributions 
results in  Eq.~\eqref{totalcurr}. For completeness
\begin{align}\label{finalc}
\frac{\delta I}{2e^2V/h} =  3&\left[(\pi T)^2
 {\color{black}+}(eV)^2\right](\Phi-\frac{2}{3}\alpha_e)(\Phi-\frac{2}{3}\alpha_o)\nonumber\\
 +&\left[(\pi T)^2 +\frac{1}{2}(eV)^2\right]\left(\alpha_e-\alpha_o\right)^2 .
\end{align}
This equation represents in a simple and transparent form contribution of the three FL constants to the charge transport.
\section{Calculation of integrals}\label{integralcomp}
\begin{figure}[t]
\includegraphics[width=6.cm, valign=c]{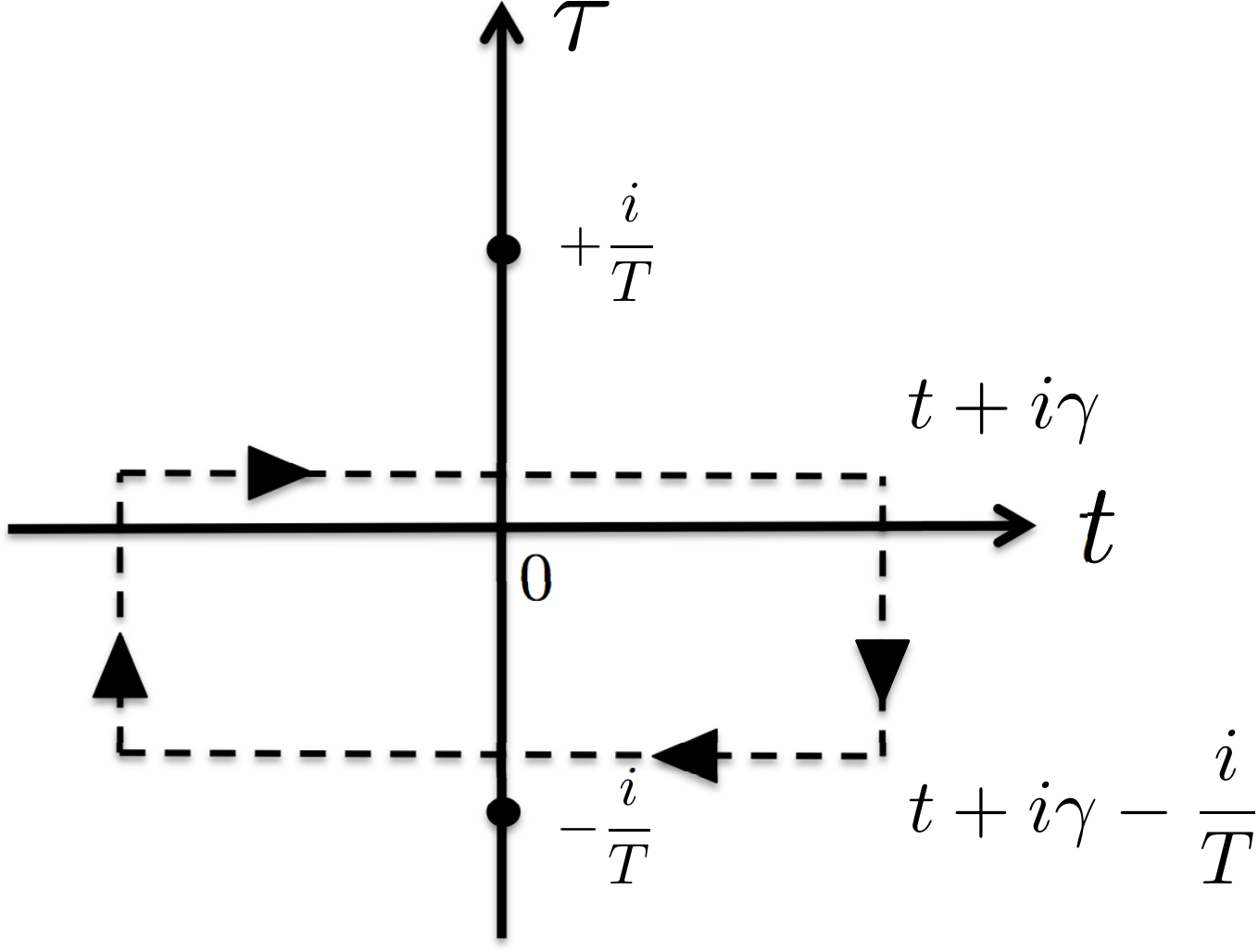}
\caption{The contour of the integration for the integral Eq.~\eqref{1a} with negative shift.}\label{cont}
\end{figure}

In this section we calculate two integrals that we used for the calculation of current correction contributed by four types of diagram. The first integral to calculate is
\begin{equation}\label{1a}
\mathcal{I}_1=\int^{\infty}_{-\infty} \frac{\cos^3(\frac{eV}{2} t) \sin(\frac{eV}{2} t)}{\sinh^4(\pi T t)} dt.
\end{equation} 
\noindent The singularity of the integral in Eq.~\eqref{1a} is removed by shifting the time contour by $i\gamma $ in the complex plane as shown in Fig.~\ref{cont}. The point splitting parameter $\gamma$
is chosen to satisfy the conditions $\gamma$$D$$\gg 1$ 
and $\gamma$$ T$$\ll 1$, $\gamma$$ e $$ V$$ \ll 1$, where, $D$ is the band cutoff. Then the Eq.\eqref{1a} can be written as
\begin{equation}\label{2a}
\begin{split}
&\mathcal{I}^+_1=\int^{\infty+i\gamma}_{-\infty+i\gamma}  \frac{\cos^3(at) \sin(at)}{\sinh^4(\pi T t)} dt\\
&{=}{-}\frac{i}{16}{\left[\mathcal{Z}(4a, T){-}\mathcal{Z}({-}4a, T)+2\mathcal{Z}(2a, T){-}2\mathcal{Z}({-}2a, T)\right]}.
\end{split}
\end{equation}
\noindent In Eq.~\eqref{2a}, $a{=}eV/2$ and we introduced the short hand notation,
\begin{equation}\label{4a}
\mathcal{Z}(a, T){=}\int^{\infty+i\gamma}_{-\infty+i\gamma}\frac{e^{iat}}{\sinh^4(\pi T t)}dt{=}
\int^{\infty+i\gamma}_{-\infty+i\gamma} h(a, T; t)dt.
\end{equation}
The poles of the integrand $h(a, T; t)$ in Eq.~\eqref{4a} are
\begin{equation}
\pi T t=\pm im\pi \Rightarrow t=\pm \frac{im}{T}, \quad m=0, \pm 1, \pm 2, \pm 3...
\end{equation}
The integration of $h(a, T; t)$ over the rectangular contour Fig.~\ref{cont} shifted by $i/T$ upon using the Cauchy residue theorem results
\begin{equation}\label{6a}
\begin{split}
\mathcal{Z}(a, T)&=\int^{\infty+i\gamma}_{-\infty+i\gamma} \frac{e^{ia(t-\frac{i}{T})}}{\sinh^4\left(\pi T(t-\frac{i}{T})\right)} dt\\
&-2\pi i\times  \left. \text{\text{Res}}[h(a, T; t)]\right|_{t=0},
\end{split}
\end{equation}
where $\text{``Res''}$ stands for the residue. By expanding the $\sinh$ function in Eq~.\eqref{6a} we get
\begin{equation}\label{7a}
\mathcal{Z}(a, T)\left(1-e^{\frac{a}{T}}\right)=-2\pi i\times  \left. \text{Res}[h(a, T; t)]\right|_{t=0}.
\end{equation}
\noindent By using the standard formula for the calculation of the residue, Eq.~\eqref{7a} cast the form
\begin{equation}\label{10a}
\mathcal{Z}(a, T)=-\frac{2\pi\left(a^3+4a(\pi T)^2\right)}{6(\pi T)^4}\times \frac{1}{1-e^{\frac{a}{T}}}.
\end{equation} 
Use of Eq.~\eqref{10a} into Eq.~\eqref{2a} gives the required integral
\begin{align}\label{fi}
\mathcal{I}^+_1&=\frac{i\pi}{(\pi T)^4}\frac{eV}{2}\left[\frac{5}{12}(eV)^2
+\frac{2}{3}(\pi T)^2\right].
\end{align}
Choosing the contour with the negative shift results in the integral $\mathcal{I}^-_1$ such that $\mathcal{I}^-_1{=}-\mathcal{I}^+_1$. As a result
\begin{align}\label{c9}
\mathcal{I}^\pm_1(V, T)&=\pm\frac{i\pi}{(\pi T)^4}\frac{eV}{2}\left[\frac{5}{12}(eV)^2
+\frac{2}{3}(\pi T)^2\right].
\end{align}
The second integral that we are going to compute is 
\begin{equation}\label{aaa}
\mathcal{I}_2=\int^{\infty}_{-\infty} \frac{\cos(\frac{eV}{2} t) \sin^3(\frac{eV}{2} t)}{\sinh^4(\pi T t)} dt.
\end{equation}
\noindent In the same way and using the same notations as for the first integral, Eq.~\eqref{aaa} reads
\begin{align}\label{bbb}
\mathcal{I}^+_2&{=}\frac{i}{16}{\left[\mathcal{Z}(4a, T){-}\mathcal{Z}({-}4a, T){-}2\mathcal{Z}(2a, T){+}2\mathcal{Z}({-}2a, T)\right]}\nonumber\\
&{=}{-}\frac{i\pi}{(\pi T)^4}\left(\frac{eV}{2}\right)^3.
\end{align}
Similar to Eq.~\eqref{c9}, the integral $\mathcal{I}_2$ takes the form
\begin{equation}\label{c10}
\mathcal{I}^\pm_2(V, T)
=\mp{\frac{i\pi}{(\pi T)^4}}{\left(\frac{eV}{2}\right)^3}.
\end{equation}
For the calculations of all diagrams we used the corresponding results of contour integration with positive shift.
%

\end{document}